\documentclass[lettersize,journal]{IEEEtran}
\usepackage{cite}
\usepackage{bm}
\usepackage{amsmath}
\usepackage{mathrsfs}

\usepackage{algorithmic}

\usepackage{array}
\usepackage{colortbl}

\ifCLASSOPTIONcompsoc
\usepackage[caption=false,font=normalsize,labelfont=sf,textfont=sf]{\part{title}subfig}
\else
\usepackage{subfigure}
\fi
\usepackage{url}
\usepackage{graphicx,amsmath,amssymb,amsfonts}
\usepackage{algorithmic,algorithm}
\usepackage{stfloats}
\usepackage{pifont}
\usepackage{amsmath}
\usepackage{makecell,rotating,multirow,diagbox}
\allowdisplaybreaks[4]
\usepackage{algorithm}
\usepackage{algorithmic}
\usepackage{setspace}

\newtheorem{lemma}{\bf Lemma}

\newtheorem{proposition}{\bf Proposition}

\def\phi{\varphi}

\def\({\left(}
\def\){\right)}

\def\b0{{\mathbf{0}}}

\usepackage{multicol}
\usepackage{xcolor}
\usepackage{multirow}
\usepackage{threeparttable}
\usepackage{framed} 
\usepackage{cancel}
\usepackage{booktabs}
\usepackage{tabularx,makecell,multirow}

\makeatletter

\newcommand{\Rmnum}[1]{\expandafter\@slowromancap\romannumeral #1@}
\newcommand{\tabincell}[2]{\begin{tabular}{@{}#1@{}}#2\end{tabular}}

\makeatother

\begin{document}
	\title{{\fontsize{21 pt}{\baselineskip}\selectfont Performance Analysis and Optimization of Reconfigurable Multi-Functional Surface Assisted Wireless Communications}}
	\author{
                 Wen~Wang,
                 Wanli~Ni,~\IEEEmembership{Graduate~Student~Member,~IEEE,}
                 Hui~Tian,~\IEEEmembership{Senior~Member,~IEEE,}
                and Naofal~Al-Dhahir,~\IEEEmembership{Fellow,~IEEE}
                \thanks{The work of Hui Tian was supported by the National Key R\&D Program of China under Grant No. 2020YFB1807801. The work of Wen Wang was supported by the Beijing University of Posts and Telecommunications (BUPT) Excellent Ph.D. Students Foundation under Grant CX2022103, and the China Scholarship Council. The work of Naofal Al-Dhahir was supported by Erik Jonsson distinguished professorship at UT-Dallas. This work was presented in part at the IEEE Global Communications Conference (GLOBECOM), Rio de Janeiro, Brazil, Dec. 2022, pp. 1-6, doi: 10.1109/GLOBECOM48099.2022.10000917. \emph{(Corresponding author: Hui Tian.)}}
				\thanks{W. Wang, W. Ni, and H. Tian are with the State Key Laboratory of Networking and Switching Technology, Beijing University of Posts and Telecommunications, Beijing 100876, China 
					(e-mail: \{wen.wang, charleswall, tianhui\}@bupt.edu.cn).}
				\thanks{N. Al-Dhahir is with the Department of Electrical and Computer Engineering, The University of Texas at Dallas, Richardson, TX 75080, USA (e-mail: aldhahir@utdallas.edu).}
	}
	\maketitle
	\begin{abstract}
		Although reconfigurable intelligent surfaces (RISs) can improve the performance of wireless networks by smartly reconfiguring the radio environment, existing passive RISs face two key challenges, i.e., double-fading attenuation and dependence on grid/battery.
	    To address these challenges, this paper proposes a new RIS architecture, called multi-functional RIS (MF-RIS).
	    Different from conventional reflecting-only RIS, the proposed MF-RIS is capable of supporting multiple functions with one surface, including signal reflection, amplification, and energy harvesting.
       As such, our MF-RIS is able to overcome the double-fading attenuation by harvesting energy from incident signals.
		Through theoretical analysis, we derive the achievable capacity of an MF-RIS-aided communication network.
		Compared to the capacity achieved by the existing self-sustainable RIS, we derive the number of reflective elements required for MF-RIS to outperform self-sustainable RIS.
		To realize a self-sustainable communication system, we investigate the use of MF-RIS in improving the sum-rate of multi-user wireless networks. 
		Specifically, we solve a non-convex optimization problem by jointly designing the transmit beamforming and MF-RIS coefficients.
		As an extension, we investigate a resource allocation problem in a practical scenario with imperfect channel state information.
		By approximating the semi-infinite constraints with the $\mathcal{S}$-procedure and the general sign-definiteness, we propose a robust beamforming scheme to combat the inevitable channel estimation errors.
		Finally, numerical results show that:
		1) compared to the self-sustainable RIS, MF-RIS can strike a better balance between energy self-sustainability and throughput improvement;
		and 2) unlike reflecting-only RIS which can be deployed near the transmitter or receiver, MF-RIS should be deployed closer to the transmitter for higher spectrum efficiency.	
	\end{abstract}
   \vspace{-7mm}
	\section{Introduction}
	
	Owing to the recent breakthroughs of metasurfaces, reconfigurable intelligent surfaces (RISs) are a promising paradigm to shape a smart radio environment for various emerging applications, such as intelligent factory and mobile holography\cite{RIS-review-Wu}.
  	Generally, RIS is a planar surface composed of a large number of energy-efficient and cost-effective passive elements, each of which induces an independent phase change to incident signals via an embedded micro-controller chip.
	Through judiciously tuning the phases of incident signals, the reradiated signals from different links are added constructively to enhance the desired reception at intended users\cite{RIS-review-Mag}, or combined destructively to alleviate information leakage at malicious eavesdroppers\cite{RIS_review_Liu}.
    Such programmable characteristics position RISs as a key enabler for throughput improvement\cite{Throughput_NOMA_TWC,Throughput_NOMA_MU-TWC}, energy reduction\cite{Throughput_NOMA_TCOM,Energy_Zhao_TCOM,Energy_Huang_TWC}, and security enhancement\cite{Secure_Li-TCOM,Secure_Wang_WCL}.

   Despite the attractive channel-modification capabilities of existing RISs, they also face significant challenges when deployed into practical systems.
   One example is the double-fading attenuation introduced by the cascaded channel, i.e., the signals relayed by the RIS experience large-scale fading twice: from the transmitter to the RIS, and  from the RIS to the receiver\cite{Active-RIS-SPM,Active-TCOM-6G}.
   This effect leads to considerable power loss when the direct link is blocked, which limits the achievable performance of RISs significantly\cite{Active-RIS-TVT-secure}.
   Although increasing the RIS size alleviates this issue, the performance gain gradually saturates when the number of passive elements exceeds a certain value\cite{Active-RIS-TWC-EE}.
   In addition, the large-size RIS usually increases production costs and deployment difficulties.
   Consequently, it is necessary to design a cost-efficient and easy-to-deploy RIS architecture to    overcome the double-fading dilemma faced by existing RIS types.
   
   Furthermore, most of the existing literature assumes that RIS is an ideal passive surface with negligible energy consumption\cite{Throughput_NOMA_TWC,Throughput_NOMA_MU-TWC}.
   However, the operation of RIS requires advanced signal processing, intelligent computation, and active electronic components, such as positive-intrinsic-negative (PIN) diodes, radio frequency (RF) switches, and phase shifters\cite{RIS-review-Wu,RIS-review-Mag}.
   These components consume a lot of energy, especially for large-size RISs with high resolution.
   For example, the operational power of RIS with $5$-bit phase shifters and $200$ elements is up to $1.2$ W, which is comparable to its energy supply and cannot be ignored\cite{Throughput_NOMA_TCOM,Energy_Zhao_TCOM,Energy_Huang_TWC}.
   Considering the non-negligible power consumption of RIS elements, it is important to develop an efficient power supply strategy to support their long-term operation.
However, existing RISs are typically non-rechargeable, making it difficult for them to achieve self-sustainability by getting rid of the dependence on battery or grid\cite{RIS_review_Liu}.
To this end, empowering RIS with energy harvesting capabilities is a promising candidate to prolong the lifetime of reflectors.
This also enhances the flexibility when selecting the deployment location of RISs as no dedicated power cables are required.
Given the ever-growing service requirements in wireless networks and the limitations of existing RISs, it is imperative to develop a new RIS architecture that is capable of mitigating the double-fading effect and achieving self-sustainability simultaneously.
	
    \begin{figure*}[t]
   	\centering
   	\includegraphics[width=6.9 in]{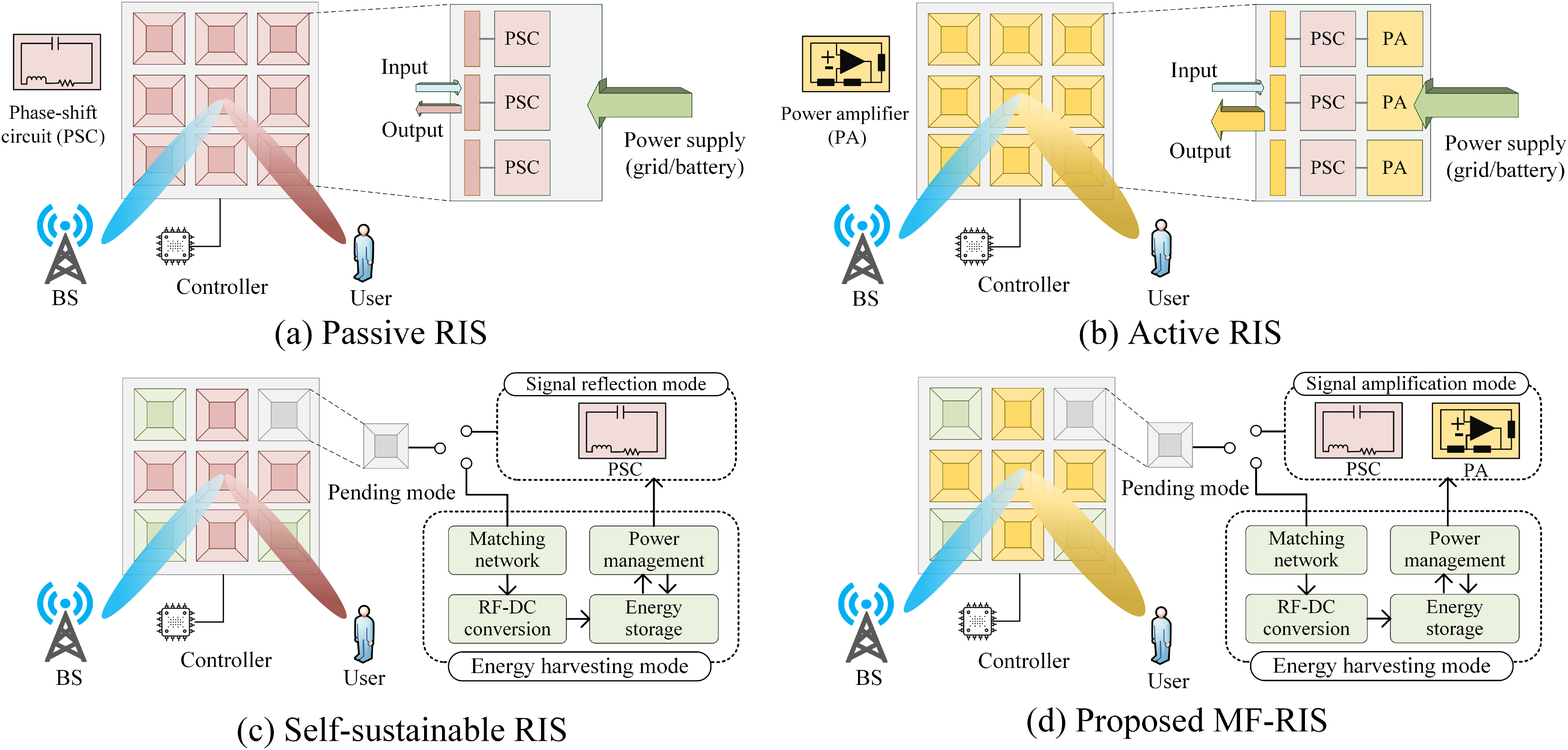}
  	\vspace{-3mm}
   	\caption{Comparison of the proposed MF-RIS with existing RIS architectures.}
   	\label{RIS_architectures}
   	\vspace{-3mm}
   \end{figure*}
\vspace{-4mm}
	\subsection{Related Works}
	In the following, we review the state-of-the-art works from two perspectives: RISs for double-fading attenuation mitigation and energy self-sustainability.
	
   \subsubsection{RISs for double-fading attenuation mitigation} 
   Considering that the conventional passive RIS suffers from the double-fading attenuation, a new RIS architecture called active RIS was proposed in \cite{Active-TCOM-6G,Active-RIS-SPM,Active-RIS-TWC-EE,Active-RIS-TVT-secure} to overcome this issue.
   In contrast to the passive RIS in Fig. \ref{RIS_architectures}(a), the active RIS in Fig. \ref{RIS_architectures}(b) activates its elements by connecting them with power amplifiers.
   As such, these elements become full-duplex active reflecting elements, allowing active RIS to amplify the power of the incident signal while modifying the phase shift.
   The authors of \cite{Active-RIS-SPM} and \cite{Active-TCOM-6G} developed a theoretical framework to compare the achievable capacity of passive and active RIS-assisted systems, and proved that active RIS can transform the multiplicative path loss into additive form.
   Motivated by this, the works \cite{Active-RIS-TVT-secure} and\cite{Active-RIS-TWC-EE} investigated the key benefits of active RIS-assisted networks in terms of secrecy improvement and throughput maximization.
   In addition, the authors of \cite{HRIS-TWC} proposed another RIS architecture enabling simultaneous signal reflection and amplification, termed hybrid relay-RIS (HR-RIS).
   Different from the active RIS that enables signal amplification by embedding negative resistance components into each element, HR-RIS requires expensive and power-hungry RF chains to amplify signals.
   This requirement leaves the practical implementation of such hybrid RIS architecture as an open problem.
   Furthermore, whether passive RISs, active RISs, or HR-RISs, they all need to be connected with a stable power supply to maintain their reflection and/or amplification circuits.
    This means that the implementation of these RISs relies on external grid or internal battery, which makes it difficult to provide flexible and uninterrupted communication services at a low cost.

   \subsubsection{RISs for energy self-sustainability}
   To eliminate the dependence on grid/battery for these conventional RISs, recent advances in RF-based energy harvesting have spawned several self-sustainable RIS architectures \cite{sustainable-RIS-TC-WPT,sustainable-RIS-TVT-WPT,sustainable-RIS-TCOM}.
   Specifically, a self-sustainable RIS enabled by wireless power transfer was proposed in \cite{sustainable-RIS-TCOM}.
   As depicted in Fig .\ref{RIS_architectures}(c), this self-sustainable RIS allows a portion of the elements to operate in signal reflection mode (R mode) to tune the wireless channels, while the remaining elements work in energy harvesting mode (H mode) to support its own operation.
   In particular, for the H mode, the incident RF signal is converted to direct current (DC) power by the energy harvesting circuit shown in Fig .\ref{RIS_architectures}(c).
   By exploiting the large number of RIS elements, such RIS simultaneously realizes self-sustainability and capacity growth.
  The authors of \cite{sustainable-RIS-TC-WPT} and \cite{sustainable-RIS-TVT-WPT} proposed another type of self-sustainable RIS, i.e., a two-phase self-sustainable RIS.
  The controller of this self-sustainable RIS can schedule its elements to working in one mode during a time slot. 
  However, the ideal linear energy harvesting model adopted in \cite{sustainable-RIS-TC-WPT} and \cite{sustainable-RIS-TVT-WPT} cannot capture the non-linear features of practical energy harvesting circuits. 

  In Table \ref{Comparison}, we compare our work with these representative works \cite{Throughput_NOMA_TCOM,Throughput_NOMA_MU-TWC,Throughput_NOMA_TWC,Energy_Zhao_TCOM,Energy_Huang_TWC,Secure_Li-TCOM,Secure_Wang_WCL,Active-TCOM-6G,Active-RIS-SPM,Active-RIS-TWC-EE,Active-RIS-TVT-secure,sustainable-RIS-TC-WPT,sustainable-RIS-TCOM,sustainable-RIS-TVT-WPT} in terms of key features, design objectives, and channel state information (CSI) setups.
  It is observed that although some recent efforts have been devoted to proposing new RIS types, there is no general RIS architecture that can simultaneously address the double-fading attenuation and the grid/battery dependence issues faced by conventional RISs.
  There are also few works to evaluate the achievable performance of RIS architectures from both optimization and analysis perspectives.
  Moreover, a robust beamforming design under imperfect CSI is still missing.
  This motivates us to propose a new RIS architecture and explore its application in practical networks.

   	\begin{table*}[t]
   	\centering
   	\renewcommand{\arraystretch}{1.1}
   	\caption{Comparison of this work with other representative works}
   	 \vspace{-2mm}
   	\label{Comparison}
   	\scalebox{0.94}{
   		\begin{tabular}{|c|c|c|c|c|c|c|c|c|c|c|c|}
   			\hline  
   			 \diagbox{Properties}{References} & \cite{Throughput_NOMA_TCOM,Throughput_NOMA_MU-TWC,Throughput_NOMA_TWC,Energy_Zhao_TCOM,Energy_Huang_TWC} & \cite{Secure_Li-TCOM}& \cite{Secure_Wang_WCL}&  \cite{Active-RIS-SPM}  &\cite{Active-RIS-TVT-secure} & \cite{Active-RIS-TWC-EE,Active-TCOM-6G} & \cite{sustainable-RIS-TCOM} & \cite{sustainable-RIS-TC-WPT,sustainable-RIS-TVT-WPT} & \textbf{This work}  \\
   		    \hline 
   		     Double-fading mitigation  & && & $\surd$  &  $\surd$  &  $\surd$  & &  & $\boldsymbol{\surd}$ \\
   		    \hline
   		     Energy self-sustainability    & && & & &  & $\surd$  & $\surd$    & $\boldsymbol{\surd}$ \\
   		     \hline
   		    Capacity analysis  &  & & $\surd$ & $\surd$ & & $\surd$  & &  & $\boldsymbol{\surd}$ \\
   		    \hline
   		    Performance optimization    & $\surd$ & $\surd$ &  & & $\surd$  & $\surd$ & $\surd$  & $\surd$   & $\boldsymbol{\surd}$ \\
   		    \hline
             Perfect CSI & $\surd$  & $\surd$ &  & $\surd$  & $\surd$  & $\surd$ & & $\surd$   & $\boldsymbol{\surd}$ \\
            \hline
            Imperfect CSI   &  & & $\surd$ &  & & & $\surd$   & & $\boldsymbol{\surd}$ \\
            \hline
             RIS architecture & \multicolumn{3}{c|}{\textbf{Passive RIS}} & \multicolumn{3}{c|}{\textbf{Active RIS}}  & \multicolumn{2}{c|}{\textbf{Self-sustainable RIS}} & \textbf{MF-RIS} \\
           \hline
   	\end{tabular}}
   \end{table*}
    
	\subsection{Motivations and Contributions}
	In this paper, we propose the multi-functional RIS (MF-RIS) as a novel RIS architecture to overcome the double attenuation and self-sustainability issues of existing RISs.
	As shown in Fig. \ref{RIS_architectures}(d), each MF-RIS element can switch between the H mode and signal amplification mode (A mode)\footnote{Existing theoretical research and prototype design of passive RISs \cite{RIS-review-Wu,RIS-review-Mag,RIS_review_Liu,Prototype-Passive-RIS}, active RISs \cite{Active-RIS-TWC-EE,Active-TCOM-6G,Active-RIS-TVT-secure,Active-RIS-SPM}, and self-sustainable RISs\cite{sustainable-RIS-TC-WPT,sustainable-RIS-TVT-WPT,sustainable-RIS-TCOM} provide a solid foundation for the implementation of the proposed MF-RIS.}.
	Specifically, the elements operating in H mode collect the RF energy from incident signals via embedded energy harvesting modules\footnote{Recent advances in multi-level converters, thin-film capacitors, and integrated power managements have greatly reduced the implementation costs of energy harvesting circuits and improved the harvesting efficiency\cite{Multi-level,Thin,Power-management}.}. 
	Meanwhile, with the help of power amplifiers and phase shift circuits, the elements in A mode reflect the incident signals with power amplification\footnote{Tunnel diode-based amplifiers enable the MF-RIS to realize signal amplification in a lightweight manner without the presence of the power-hungry RF chain components\cite{Active-hardware-1}.}.
	Equipped with the capability of joint signal reflection, amplification, and energy harvesting, the proposed MF-RIS facilitates self-sustainability while maintaining performance advantages.
	Our main contributions are summarized as follows:
	\begin{itemize}
		\item We propose a new MF-RIS architecture enabling simultaneous signal reflection, amplification, and energy harvesting. 
	   Specifically, we provide the physical implementation and operating protocol of  MF-RIS from the wireless communications perspective.
	Then, we analyze the achievable capacity performance of MF-RIS and compare it with self-sustainable RIS.
		Our theoretical results reveal that the proposed MF-RIS outperforms the self-sustainable RIS in terms of achievable signal-to-noise ratio (SNR).
		\item 
		We formulate a sum-rate (SR) maximization problem for an MF-RIS-aided multi-user system, and solve the resulting mixed-integer non-linear programming (MINLP) problem by optimizing the transmit beamforming and MF-RIS coefficients iteratively. 
       Considering the inevitable CSI estimation error, we propose a robust beamforming scheme to maximize the SR of all users.
       Specifically, to handle the infinite possibilities introduced by CSI uncertainties, we adopt the $\mathcal{S}$-procedure and the general sign-definiteness to approximate semi-infinite constraints and convert them into finite ones.
		\item Simulation results are provided to verify the effectiveness and robustness of the proposed algorithms. 
		In particular, the following observations are made from extensive numerical results:
		1) compared to the self-sustainable RIS, the MF-RIS can attain 114\% higher SR gain, by integrating multiple functions on one surface;
		and 2) increasing the number of elements is beneficial for an improved self-sustainability and throughput, but this also amplifies the performance loss caused by imperfect CSI, especially when the channel uncertainty is high.
	\end{itemize}

	The rest of this paper is organized as follows. 
	In Section \ref{Syetem_Model}, we provide the system model of an MF-RIS-aided communication network.
   In Section \ref{Analysis results}, we analyze the achievable capacity of the proposed MF-RIS.
	In Section \ref{Perfect CSI case}, we formulate a SR maximization problem, which is solved by an iterative algorithm.
    In Section \ref{Imperfect CSI case}, we extend the problem and algorithm to the imperfect CSI case.
	Numerical results are presented in Section \ref{Numerical Results}, followed by conclusions in Section \ref{Conclusion}.
	
	\emph{Notations:} 
	$\mathbb{H}^{N}$ denotes the complex Hermitian matrix with $N\times N$ dimensions.
	$\mathbf{X}^{\ast}$, $\mathbf{X}^{\rm T}$, $\mathbf{X}^{\rm H}$, $\lVert\mathbf{X} \lVert_F$, and ${\rm vec}(\mathbf{X})$ denote the conjugate, transpose, Hermitian, Frobenius norm, and vectorization of matrix $\mathbf{X}$, respectively.
	$\lVert\mathbf{x} \lVert$ denotes the Euclidean norm of vector $\mathbf{x}$.
	${\rm Re}\{\cdot\}$ denotes the real part of a complex number.
	${\rm diag}(\cdot)$, $\operatorname{mod}$, $\lceil \cdot \rceil$, and $\lfloor \cdot \rfloor$ denote the diagonal operation, the modulus operation, the rounding up and rounding down operations, respectively.
	$[\mathbf{X}]_{m,m}$ and $[\mathbf{x}]_{m}$ denote the $m$-th diagonal element and the $m$-th element of matrix $\mathbf{X}$ and vector $\mathbf{x}$, respectively.
	$\otimes$ and $\odot$ denote the Kronecker product and the Hadamard product, respectively.
	$\mathbf{1}_M$ is an $M\times 1$ all-ones vector. $\mathbf{X}\succeq \mathbf{0}$ indicates that matrix $\mathbf{X}$ is positive semi-definite.

	\begin{figure}[t]
	\centering
	\includegraphics[width=3.4 in]{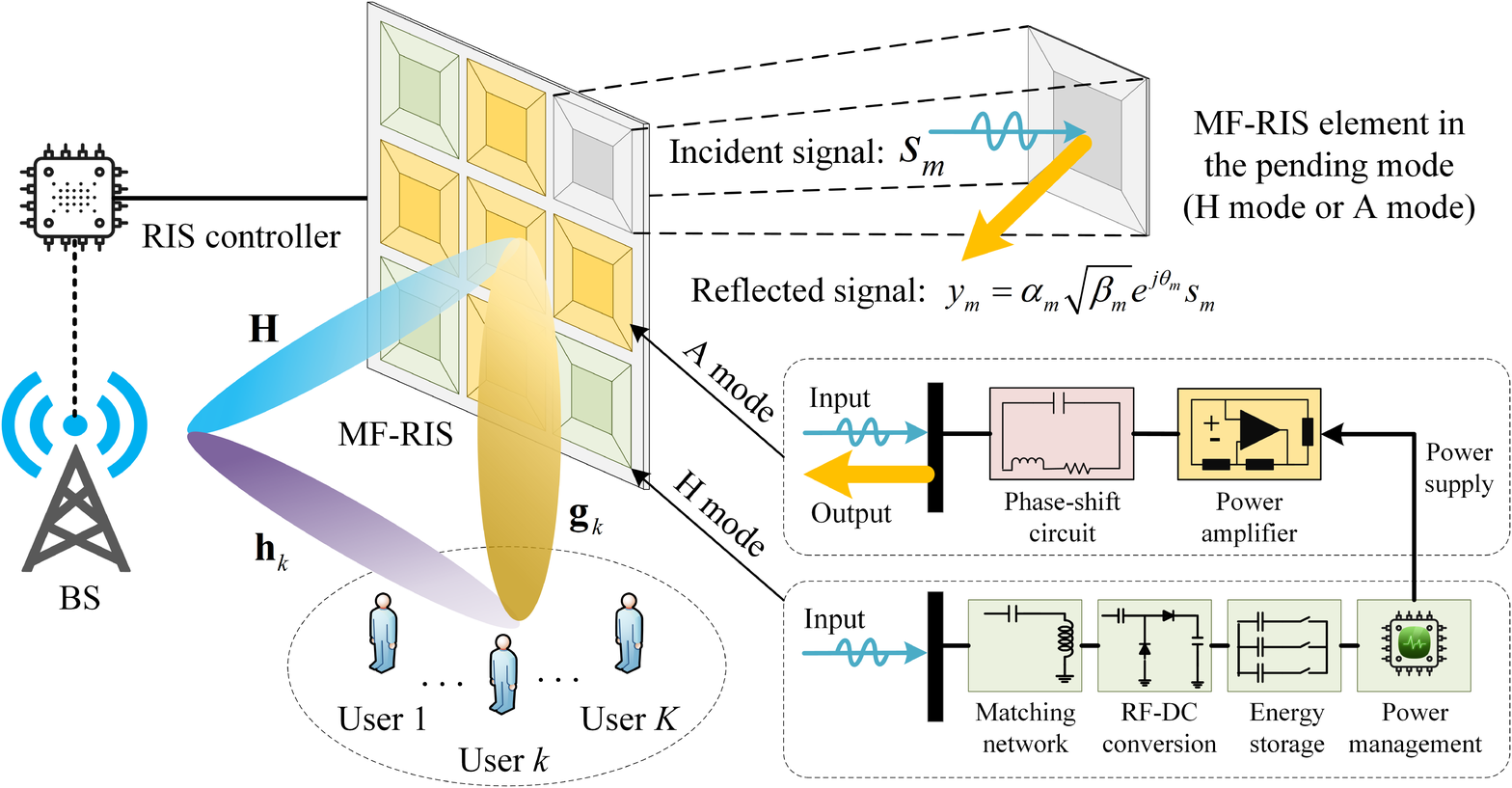}
	\vspace{-2mm}
	\caption{An MF-RIS-aided downlink multi-user communication system.}
	\label{system_model}
    \end{figure}
   
   \vspace{-2mm}
	\section{System Model}\label{Syetem_Model}
	As shown in Fig. \ref{system_model}, we consider an MF-RIS-assisted multi-user downlink communication network, where an MF-RIS is deployed to assist wireless communications from an $N$-antenna base station (BS) to $K$ single-antenna users.
	The set of users is denoted by $\mathcal{K}=\{1,2,\cdots,K\}$.
	We assume that the MF-RIS is equipped with $M$ elements, denoted by $\mathcal{M}=\{1,2,\cdots,M\}$.
	These $M$ elements are divided into two groups, with one operating in H mode and the other in A mode.
	Specifically, the group of elements in H mode harvests the RF energy from received signals to support MF-RIS operation.
	Meanwhile, the remaining elements in A mode reflect and amplify the incident signals.
	The MF-RIS coefficient matrix is denoted by
	$\boldsymbol{\Theta}={\rm{diag}}(\alpha_1\sqrt{\beta_1}e^{j\theta_1},\alpha_2\sqrt{\beta_2}e^{j\theta_2},\cdots,\alpha_M\sqrt{\beta_M}e^{j\theta_M})$,
	where $\alpha_m\in\{0,1\}$, $\beta_m\in[0,\beta_{\rm max}]$, and $\theta_m \in[0,2\pi)$ denote the mode indicator, amplitude, and phase shift of the $m$-th element, respectively.
	Here, $\alpha_m=1$ implies that the $m$-th element operates in A mode, while $\alpha_m=0$ implies that it works in H mode, and $\beta_{\rm max}\geq 1$ represents the amplification factor.
	Therefore, by optimizing the mode indicator $\alpha_{m}$, the $M$ elements are flexibly assigned to work in A or H mode.
	
	Denote $\mathbf{w}=\sum_{k=1}^{K}\mathbf{f}_ks_{k}$ as the superposition signal transmitted by the BS, where $\mathbf{f}_k$ is the transmit beamforming vector of user $k$, and $s_k\sim\mathcal{CN}(0,1)$ denotes the modulated data symbol, which is independent over $k$.
	The signal received at user $k$ is then given by
	\setlength{\abovedisplayskip}{3pt}
	\setlength{\belowdisplayskip}{3pt}
	\begin{align}
		y_k=\left(\mathbf{h}_{k}^{\rm H}+\mathbf{g}_{k}^{\rm H}\boldsymbol{\Theta}\mathbf{H}\right)\mathbf{w}+\mathbf{g}_{k}^{\rm H}\boldsymbol{\Theta}\mathbf{n}_s+n_k, 
	\end{align}
	where $\mathbf{n}_s\sim \mathcal{C} \mathcal{N}(\mathbf{0}, \sigma_1^{2}\mathbf{I}_M)$ denotes the thermal noise generated at the MF-RIS with per-element noise power $\sigma_1^2$, and $n_{k} \sim \mathcal{C} \mathcal{N}(0, \sigma_0^{2})$ denotes the additive white Gaussian noise (AWGN) at user $k$ with noise power $\sigma_0^{2}$.
	In addition,
	$\mathbf{h}_{k}^{\rm H}\in\mathbb{C}^{1\times N}$,  $\mathbf{H}\in\mathbb{C}^{M\times N}$, and $\mathbf{g}_{k}^{\rm H}\in\mathbb{C}^{1\times M}$ represent the channels from the BS to user $k$, from the BS to the MF-RIS, and from the MF-RIS to user $k$, respectively.
	Accordingly, by defining $\bar{\mathbf{h}}_k=\mathbf{h}_{k}^{\rm H}+\mathbf{g}_{k}^{\rm H}\boldsymbol{\Theta}\mathbf{H}$ as the combined channel from the BS to user $k$, the achievable  rate of user $k$ is given by
	\begin{eqnarray}
		R_k=\log_2\left(1+\frac{|\bar{\mathbf{h}}_k\mathbf{f}_k|^2}{\sum\nolimits_{i=1,i\neq k}^{K}|\bar{\mathbf{h}}_k\mathbf{f}_i|^2 +\sigma_1^2\lVert\mathbf{g}_{k}^{\rm H}\boldsymbol{\Theta}\lVert^2+\sigma_0^2}\right).
	\end{eqnarray}
	
	We define $\mathbf{T}_m={\rm diag}([\underbrace{0,\cdots,}_{1 \ \text{to} \ m-1} 1-\alpha_{m} \underbrace{,\cdots, 0}_{m+1 \ \text{to} \ M}])$ as the mode indicator matrix of the $m$-th element.
	The RF power received at the $m$-th element is then expressed as
	\begin{eqnarray}
		\label{MF-RIS-RF}
		P_m^{\rm{RF}}=\mathbb{E}\left(\left\| \mathbf{T}_m\left(\mathbf{H}\mathbf{w}+\mathbf{n}_{s}\right)\right\|^{2}\right), 
	\end{eqnarray}
	where the expectation operator $\mathbb{E}(\cdot)$ is taken over $\mathbf{w}$ and $\mathbf{n}_s$.
	Based on the logistic function\cite{WCL-Nonlinear-cost reduction}, we adopt a non-linear energy harvesting model for the MF-RIS.
	Specifically, the total power harvested at the $m$-th element is modeled as
	\begin{eqnarray}
		\label{Energy_harvesting_model}
		P_m^{\mathrm{A}}=\frac{\Upsilon_m-Z\Omega}{1-\Omega},~~\Upsilon_m=\frac{Z}{1+e^{-a (P_m^{\rm RF}-q)}}, 
	\end{eqnarray}
	where $\Upsilon_m$ is a logistic function with respect to the received RF power $P_{m}^{\rm RF}$, and $Z$ is a constant determining the maximum harvested power.
	Here, constant $\Omega$ is given by $\Omega=\frac{1}{1+e^{aq}}$, $a>0$ and $q>0$ are related to circuit properties such as the capacitance and diode turn-on voltage\cite{2015-CL-energy harvesting}.
	
	To achieve the self-sustainability of MF-RIS, the total power consumed at the MF-RIS should not exceed its harvested power, given by\cite{sustainable-RIS-TCOM}
	\begin{align}
		\nonumber
		&\sum\nolimits_{m=1}^{M} \alpha_{m}(P_{b}  + P_{\rm DC})+(M-\sum\nolimits_{m=1}^{M}\alpha_{m})P_{\rm C} \\
		\label{C_energy}
		&+ \xi P_{\rm O} \leq \sum\nolimits_{m=1}^{M} P_m^{\rm A}, 
	\end{align}
	where $P_{b}$, $P_{\rm DC}$, and $P_{\rm C}$ denote the amount of power consumed by each phase shifter, the DC biasing power consumption of the amplifier circuit, and the power consumption of the RF-to-DC power conversion circuit, respectively\cite{Active-RIS-TWC-EE}.
	Here, $\xi$ is the inverse of the amplifier efficiency, and $P_{\rm O }=\sum\nolimits_{k=1}^{K}\lVert \boldsymbol{\Theta}\mathbf{H}\mathbf{f}_k\lVert^2+\sigma_1^2\lVert \boldsymbol{\Theta}\lVert_{F}^2$ is the output power of MF-RIS.
     
     In this paper, we focus on the throughput analysis and optimization of the considered MF-RIS-aided communication network.
     To illustrate the superiority of the proposed MF-RIS architecture, we first analyze the achievable capacity in Section \ref{Analysis results}.
     Then, we maximize the achievable SR under perfect and imperfect CSI cases in Sections \ref{Perfect CSI case} and \ref{Imperfect CSI case}, respectively.
	
	\vspace{-1mm}
	\section{Capacity Comparison between MF-RIS and Self-sustainable RIS} \label{Analysis results}
	To gain more insights, we analyze the performance gain achieved by the MF-RIS in a simplified single-input single-output (SISO) system with one user.
	Focusing on the capacity of MF-RIS-aided channels, the direct link is assumed to be blocked, and the reflection link is line-of-sight (LoS)\footnote{
	To better understand the characteristics of the proposed MF-RIS, we consider the tractable case of the SISO system and the LoS channel, similar to 	\cite{Active-RIS-TWC-EE,Active-CL-Full,Active-WCL-2022,Active-WCL-You}.
    The numerical results in Section \ref{Numerical Results} verify that the LoS case characterizes a performance upper bound for the MF-RIS-aided system, and the corresponding insights can be used to guide the more general cases.
	}.
	Given the mode indicator matrix $\boldsymbol{\alpha}={\rm{diag}}(\alpha_1,\alpha_2,\cdots,\alpha_M)$, the signal received at the user is given by
	\begin{align}
		y=\mathbf{g}^{\rm H}\boldsymbol{\alpha}\boldsymbol{\Xi}\mathbf{h}\sqrt{p}+\mathbf{g}^{\rm H}\boldsymbol{\alpha}\boldsymbol{\Xi}\mathbf{n}_s+n,
	\end{align}
	where $p$ denotes the transmit power at the BS, $\boldsymbol{\Xi}\!=\!\operatorname{diag}(\sqrt{\beta_1}e^{j\theta_1},\sqrt{\beta_2}e^{j\theta_2},\cdots,\sqrt{\beta_M}e^{j\theta_M})$, and $n\!\sim\!\mathcal{CN}(0,\sigma_0^2)$ denotes the AWGN noise at the user.
	Moreover, $\mathbf{g}=g\mathbf{a}(\phi_a,\phi_e)$ and $\mathbf{h}\!=\!h\mathbf{a}(\psi_a,\psi_e)$ represent the channel vectors from the MF-RIS to the user and from the BS to the MF-RIS, respectively, where $g$ and $h$ denote the distance-dependent path-loss factors, $\phi_a$, $\phi_e$, $\psi_a$, and $\psi_a$ denote the azimuth and elevation angles of arrival and departure, respectively. Here, $\mathbf{a}(\psi_a,\psi_e)$ and $\mathbf{a}(\phi_a,\phi_e)$ denote steering vector functions with respect to these angles\cite{Active-WCL-2022}.
	Then, the SNR maximization problem is formulated as
	\begin{subequations}
		\label{analysis-problem-1}
		\begin{eqnarray}
			\label{Function-SNR-analysis}
			&\underset{p,\boldsymbol{\Xi}}{\max}  &\frac{p|\mathbf{g}^{\rm H}\boldsymbol{\alpha}\boldsymbol{\Xi}\mathbf{h}|^2}
			{\sigma_1^2\left\|\mathbf{g}^{\rm H}\boldsymbol{\alpha}\boldsymbol{\Xi}\right\|^2+\sigma_0^2}\\
			&\operatorname{s.t.} &  p\in [0, P_{\rm BS}^{\rm max}],\\
			\label{analysis-problem-1-C1}
			&&\beta_m\in [0,\beta_{\max}], ~\theta_m\in[0,2\pi), ~\forall m,\\
			\label{analysis-problem-1-C2}
			&&p\lVert\boldsymbol{\alpha}\boldsymbol{\Xi}\mathbf{h}\lVert^2+\sigma_1^2\lVert\boldsymbol{\alpha}\boldsymbol{\Xi}\lVert_F^2\leq P_{\rm O}^{\rm MF}(\boldsymbol{\alpha}),
		\end{eqnarray}
	\end{subequations}
	where $P_{\rm O}^{\rm MF}(\boldsymbol{\alpha})=\frac{1}{\xi}(\sum\nolimits_{m=1}^M P_m^{\rm A}-\sum\nolimits_{m=1}^M\alpha_m(P_b+P_{\rm DC})-(M-\sum\nolimits_{m=1}^M\alpha_m)P_{\rm C})$ denotes the maximum output power of the MF-RIS.
	By exploiting the Lagrangian duality, the optimal transmit power and phase shift for problem (\ref{analysis-problem-1}) are, respectively, obtained as\cite{Active-CL-Full}
	\begin{align}
		\label{C_analysis-problem_solution1}
	   p^{\star}\!=\!P_{\rm BS}^{\rm max},
		~\theta_m^{\star}\!=\!(\operatorname{arg}\{ [\mathbf{g}]_m \} \! - \!\operatorname{arg}\{ [\mathbf{h}]_m \}) \bmod 2\pi, \forall m.\!\!\!
	\end{align}
	\begin{proposition}
		\label{proposition-1}
		\emph{Assume that the numbers of  MF-RIS elements operating in A mode and H mode are $M_{\rm A}=\sum_{m=1}^M \alpha_{m}$ and $M_{\rm H}=M-\sum_{m=1}^M \alpha_{m}$, respectively.
		Then, the optimal magnitude coefficient for problem (\ref{analysis-problem-1}) is given by
			\begin{align}
				\label{C_analysis-problem_solution2}		
				\beta^{\star}_m=\begin{cases}\beta_{\rm max},  & M_{\rm A}\leq  M_{\rm A,1}, \\ 
					\frac{P_{\rm O}^{\rm MF}(\boldsymbol{\alpha})}{ M_{\rm A}(P_{\rm BS}^{\max}h^2+\sigma_1^2)}, & M_{\rm A}> M_{\rm A,1},
				\end{cases}
			\end{align}
	where
		\begin{align}
			M_{\rm A,1}=\frac{\sum\nolimits_{m=1}^M P_m^{\rm A}-M_{\rm H}P_{\rm C}}{ \xi \beta_{\rm max}(P_{\rm BS}^{\max}h^2+\sigma_1^2)+P_b+P_{\rm DC}}. 
		\end{align}
		}
	\end{proposition}
	\begin{IEEEproof}
		Please refer to Appendix \ref{proof_of_proposition_1}.
	\end{IEEEproof}
	
	\begin{proposition}
		\label{MF-RIS-SNR}
		\emph{
			 Based on the optimal solutions in (\ref{C_analysis-problem_solution1}) and (\ref{C_analysis-problem_solution2}), the achievable SNR achieved by the MF-RIS is given by
			 \setlength{\abovedisplayskip}{5pt}
			 \setlength{\belowdisplayskip}{5pt}
			\begin{eqnarray}
				\label{MF-RIS-SNR-2}
				\gamma_{{\rm MF}}\!\!\!\!\!\!\!
				&&=\left\{ \begin{aligned} 
					&\frac{P_{\rm BS}^{\rm max}\beta_{\max}h^2g^2M_{\rm A}^2}{\beta_{\max}\sigma_1^2g^2M_{\rm A}+\sigma_0^2}, & & M_{\rm A}\leq M_{\rm A,1},  \\ 
					&\frac{P_{\rm BS}^{\rm max}h^2g^2P_{\rm O}^{\rm MF}(\boldsymbol{\alpha})M_{\rm A}}{\sigma_1^2g^2P_{\rm O}^{\rm MF}(\boldsymbol{\alpha})+\sigma_0^2(P_{\rm BS}^{\max}h^2+\sigma_1^2)}, &    & M_{\rm A}> M_{\rm A,1}. \\ 
				\end{aligned} \right. 
			\end{eqnarray}
		}
	\end{proposition}
	\begin{IEEEproof}
		Please refer to Appendix \ref{proof_of_MF-RIS-SNR}.
	\end{IEEEproof}
	\begin{proposition}
		\label{proposition-2}
		\emph{For the considered MF-RIS-assisted system, the optimal number of reflection elements is given by
			\setlength{\abovedisplayskip}{3pt}
			\setlength{\belowdisplayskip}{3pt}
		\begin{align}
			\label{MF-Optimal-M}
		   M_{\rm A}^{\star}= \begin{cases}
		   		\lfloor \bar{M}_{\rm A} \rfloor,  & \gamma_{{\rm MF}}( \lceil \bar{M}_{\rm A} \rceil) \leq \gamma_{{\rm MF}}( \lfloor \bar{M}_{\rm A} \rfloor),\\
		   	\lceil \bar{M}_{\rm A} \rceil,  & \gamma_{{\rm MF}}( \lceil \bar{M}_{\rm A} \rceil) > \gamma_{{\rm MF}}( \lfloor \bar{M}_{\rm A} \rfloor), 
		   \end{cases}  
		\end{align}	
	   where $\bar{M}_{\rm A}=\max\{M_{\rm A,1},M_{\rm A,2}\} $.
		Denote $\mathcal{W}_1=\sum_{m=1}^MP_m^{\rm A}-M_{\rm H}P_{\rm C}$, $\mathcal{W}_2=P_b+P_{\rm DC}$, $\mathcal{W}_3=\sigma_1^2g^2$, and $\mathcal{W}_4=\xi\sigma_0^2(P_{\rm BS}^{\max}h^2+\sigma_1^2)$.
		Then, the value of $M_{\rm A,2}$ is given by
			\begin{align}
				\label{M-2}
				M_{\rm A,2}=\frac{\mathcal{W}_1\mathcal{W}_3+\mathcal{W}_4-\sqrt{\mathcal{W}_1\mathcal{W}_3\mathcal{W}_4+\mathcal{W}_4^2}}{\mathcal{W}_2\mathcal{W}_3}.
			\end{align}}
	\end{proposition}
	
	\begin{IEEEproof}
		Please refer to Appendix \ref{proof_of_proposition_2}.
	\end{IEEEproof}
	
	Next, we analyze the achievable capacity of the self-sustainable RIS proposed in\cite{sustainable-RIS-TCOM}.
	Assume that the numbers of elements operating in H mode and R mode are $M_{\rm H}$ and $M_{\rm A}$, respectively.
	Given the mode indicator matrix $\boldsymbol{\alpha}$, the SNR maximization problem is formulated as
	\begin{subequations}
		\label{analysis-problem-STAR-1}
		\begin{eqnarray}
			\label{analysis-SE-function}
			& \underset{p,\boldsymbol{\Xi}}{\max}  &\frac{p|\mathbf{g}^{\rm H}\boldsymbol{\alpha}\boldsymbol{\Xi}\mathbf{h}|^2}{\sigma_0^2}\\
			&\operatorname{s.t.} &  p\in[0, P_{\rm BS}^{\rm max}], \\
			&&\beta_m\in [0,1], ~\theta_m\in[0,2\pi), ~\forall m,\\
			\label{power-DF-RIS}
			&&M_{\rm A}P_b+M_{\rm H}P_{\rm C}\leq \sum\nolimits_{m=1}^M P_m^{\rm A}.
		\end{eqnarray}
	\end{subequations}

	Again, using the Lagrangian duality, the optimal solution for problem (\ref{analysis-problem-STAR-1}) is obtained as \cite{Active-CL-Full}
	\begin{subequations}
		 \begin{align}
			\label{DF-RIS-solution-1}	
			&p^{\star}=P_{\rm BS}^{\max},
			~\beta^{\star}_m=1, \\
			\label{DF-RIS-solution-2}	
			&\theta_m^{\star}=(\operatorname{arg}\{ [\mathbf{g}]_m \}-\operatorname{arg}\{ [\mathbf{h}]_m \}) \bmod 2\pi, ~\forall m.
		\end{align}
	\end{subequations}
	\begin{proposition}
		\label{proposition-3}
		\emph{The achievable SNR of the considered self-sustainable RIS-aided system is 
	\begin{align}
	\label{SNR-SE-RIS}
	\gamma_{{\rm SE}}
     = \frac{P_{\rm BS}^{\rm max}h^2g^2M_{\rm A}^2}{\sigma_0^2}.
 \end{align}
			Moreover, the optimal number of elements operating in R mode is $M_{\rm A}^{\star}=\lfloor \frac{\sum_{m=1}^MP_m^{\rm A}-M_{\rm H}P_{\rm C}}{P_b}\rfloor$.
		}
	\end{proposition}
	\begin{IEEEproof}
		Please refer to Appendix \ref{proof_of_proposition_3}.
	\end{IEEEproof}
	
	According to (\ref{MF-RIS-SNR-2}) and (\ref{SNR-SE-RIS}), by solving $\gamma_{{\rm MF}}\geq\gamma_{\rm SE}$, we obtain the number of reflection elements required for MF-RIS to outperform self-sustainable RIS, which is given by
	\begin{align}
		\nonumber
		&M_{\rm A}\leq M_{\rm A}^{\rm th}\\
		\label{n-constraint}
		&=\min\Bigg\{
		\frac{(\beta_{\rm max} \! - \!1)\sigma_0^2}{\beta_{\rm max}\mathcal{W}_3}, 
		\frac{\mathcal{W}_5 \! - \! \sqrt{\mathcal{W}_5^2 \! -\! 4\mathcal{W}_1\mathcal{W}_2\mathcal{W}_3\sigma_0^2}}{2\mathcal{W}_2\mathcal{W}_3}\Bigg\},\!\!\!
	\end{align}
	where $\mathcal{W}_5=\mathcal{W}_2\sigma_0^2+\mathcal{W}_1\mathcal{W}_3+\mathcal{W}_4$.
    For a more intuitive comparison, we set $P_{\rm BS}^{\rm max}=5$ W, $M=300$,  $\sigma_0^2=\sigma_1^2=-70$ dBm, $h^2=-45$ dB, $g^2=-60$ dB,  $P_b=1.5$ mW, $P_{\rm C}=2.1\ \mu$W, $P_{\rm DC}=0.3$ mW, $\beta_{\max}=13$ dB, and $\xi=1.1$\cite{sustainable-RIS-TCOM}.
	Then, when $M_{\rm A}\leq  21$, the achievable SNR performance of the MF-RIS is better than the self-sustainable RIS counterpart, i.e., the inequality (\ref{n-constraint}) holds.
   In particular, for a practical RIS size, e.g., $M_{\rm A}=10$, we obtain $\gamma_{{\rm MF}}\!\approx \!33.2$ dB and $\gamma_{{\rm SE}}\!\approx \!22$ dB, where the former is about $13.2$ times larger than the latter.
	
	\vspace{-1mm}
	\section{Throughput Maximization Under Perfect CSI}\label{Perfect CSI case}
	In this section, we maximize the throughput in a multiple-input single-output (MISO) system where multiple users are assisted by an MF-RIS, as shown in Fig. \ref{system_model}.
	Specifically, considering the power budget and energy causality in the perfect CSI case, we propose an iterative algorithm to solve the resulting MINLP problem efficiently.
	
	\vspace{-2mm}
	\subsection{Problem Formulation Under Perfect CSI}
	Our objective is to maximize the achievable SR of all users by jointly optimizing the transmit beamforming at the BS and the MF-RIS coefficients,  while maintaining the self-sustainability of MF-RIS.
	In this section, to characterize the performance upper bound achieved by the MF-RIS, we assume that the perfect CSI of all channels is available at the BS by applying existing channel estimation methods\cite{Channel_estimation_You_JSAC}.
	Mathematically, the optimization problem is formulated as 
	\begin{subequations}
		\label{P0}
		\begin{eqnarray}
			\label{P0_function}
			& \underset{\mathbf{f}_k,\boldsymbol{\Theta}}{\max}  & \sum\nolimits_{k=1}^KR_{k}\\
			\label{C_transmit beamforming}
			& \operatorname{s.t.} &\sum\nolimits_{k=1}^{K} \lVert \mathbf{f}_k\lVert^2\leq P_{\rm BS}^{\max},\\
			\label{C-MF-RIS}
			&&\boldsymbol{\Theta}\in \mathcal{R}_{\rm MF},\\
			&&{\rm (\ref{C_energy})}.
		\end{eqnarray}
	\end{subequations}
     where $P_{\rm BS}^{\max}$ is the power budget at the BS and $\mathcal{R}_{\rm MF}\!=\!\{\alpha_m,\beta_m,\theta_m|\alpha_m\!\in\!\{0,1\},\beta_m\in[0,\beta_{\max}], $ $\theta_m\in[0,2\pi),\forall m\}$ is the feasible set of MF-RIS coefficients.
	
	Unlike conventional RIS-related works that only focus on the optimization of reflective phase shifts\cite{Throughput_NOMA_MU-TWC,Throughput_NOMA_TWC}, in this paper, we consider the joint design of mode indicator, amplitude, and phase shift coefficients of the MF-RIS.
	This results in a more challenging optimization problem, where the newly introduced signal amplification and energy harvesting functions bring more highly-coupled non-convex constraints.
	Specifically, the signal amplification introduces additional RIS noise in the objective function (\ref{P0_function}) and constraint (\ref{C_energy}), which complicates the resource allocation problem.
	Since the adopted non-linear energy harvesting model involves complex logistic functions, constraint (\ref{C_energy}) is more difficult to deal with than the linear energy constraint in \cite{sustainable-RIS-TC-WPT} and \cite{sustainable-RIS-TVT-WPT}.
	Furthermore, the proposed MF-RIS requires the joint optimization of binary mode indicators and continuous amplitude and phase shift coefficients, which introduces a mixed-integer constraint in (\ref{C_energy}).
	As a result, problem (\ref{P0}) is an MINLP problem, which is more general than the optimization problems considered in \cite{Throughput_NOMA_MU-TWC} and \cite{Throughput_NOMA_TWC}. 
	However, it cannot be solved directly due to the non-convex and non-linear features.
	In the following, we develop an alternating optimization (AO) algorithm to find a high-performance solution with low complexity.
	
	\vspace{-2mm}
	\subsection{Problem Transformation Under Perfect CSI}
	Before solving problem (\ref{P0}), we first transform it into a more tractable form.
	To tackle the non-concave objective function (\ref{P0_function}), we introduce auxiliary variables $Q_{k}$, $\mathcal{A}_{k}$, and $\mathcal{B}_{k}$, 
	satisfying $Q_{k}=\log_2\left(1+{\mathcal{A}_{k}^{-1}\mathcal{B}_{k}^{-1}}\right)$, $\mathcal{A}_{k}^{-1}=|\bar{\mathbf{h}}_{k}\mathbf{f}_{k}|^2$, and $\mathcal{B}_{k}=\sum\nolimits_{i=1,i\neq k}^{K}|\bar{\mathbf{h}}_{k}\mathbf{f}_{i}|^2+\sigma_1^2\lVert\mathbf{g}_{k}^{\mathrm H}\boldsymbol{\Theta}\lVert^2+\sigma_0^2$.
	With these variable definitions, we obtain the following new constraints:
	\begin{subequations}
		\begin{align}
			\label{C_Q}
			&Q_{k}\leq \log_2\left(1+{\mathcal{A}_{k}^{-1}\mathcal{B}_{k}^{-1}}\right), ~\forall k,\\
			\label{C_AB_1}
			&{\mathcal{A}_{k}^{-1}}\leq |\bar{\mathbf{h}}_{k}\mathbf{f}_{k}|^2, ~\forall k,\\
			\label{C_AB_2}
			& \mathcal{B}_{k}\geq \sum\nolimits_{i=1,i\neq k}^{K}|\bar{\mathbf{h}}_{k}\mathbf{f}_{i}|^2+\sigma_1^2\lVert\mathbf{g}_{k}^{\mathrm H}\boldsymbol{\Theta}\lVert^2+\sigma_0^2,~\forall k.
		\end{align} 
	\end{subequations}
     
     To handle the non-convexity of constraint (\ref{C_Q}), we here exploit the successive convex approximation (SCA) technique.
	Using the fact that the first-order Taylor expansion (FTS) of a convex function is a global under-estimator, a lower bound on its right-hand-side (RHS) at the feasible point $\{\mathcal{A}_{k}^{(\ell)},\mathcal{B}_{k}^{(\ell)}\}$ in the $\ell$-th iteration is given by
	\begin{align}
		\nonumber
		R_{k}^{\rm lb}=&\log_2\Big(1+\frac{1}{\mathcal{A}_{k}^{(\ell)}\mathcal{B}_{k}^{(\ell)}}\Big) \\
		&-\frac{(\log_2 e)(\mathcal{A}_{k}-\mathcal{A}_{k}^{(\ell)})}{\mathcal{A}_{k}^{(\ell)}+(\mathcal{A}_{k}^{(\ell)})^2\mathcal{B}_{k}^{(\ell)}}
		-\frac{(\log_2 e)(\mathcal{B}_{k}-\mathcal{B}_{k}^{(\ell)})}{\mathcal{B}_{k}^{(\ell)}+(\mathcal{B}_{k}^{(\ell)})^2\mathcal{A}_{k}^{(\ell)}}.
	\end{align} 

	To facilitate the derivation of constraint (\ref{C_energy}), we first rewrite the received RF power and the output power as follows
	\begin{subequations}
	\begin{align}
		&\!\!\!\!\!\!P_m^{\rm{RF}}\!\!=\!{\rm Tr}\Big( \mathbf{T}_m\mathbf{H}\big(\sum\nolimits_{k=1}^K\mathbf{f}_k\mathbf{f}_k^{\rm H}\big)\mathbf{H}^{\rm H}\mathbf{T}_m^{\rm H}\Big) \! + \! (1 \! -\! \alpha_{m})\sigma_1^2, \!\!\!\!\!\!\!\!\\
		&\!\!\!\!\!\!P_{\rm O}\!=\!{\rm  Tr}\Big(\mathbf{\Theta}\big(\mathbf{H} (\sum\nolimits_{k=1}^K\mathbf{f}_k\mathbf{f}_k^{\rm H})\mathbf{H}^{\rm H}+\sigma_1^2\mathbf{I}_M\big)\mathbf{\Theta}^{\rm H}\Big).
	\end{align}
    \end{subequations}
	Then, by introducing auxiliary variables $\mathcal{C}_m$ and $\zeta_m$, constraint (\ref{C_energy}) is equivalently recast as
	\begin{subequations}
		\begin{align}
			\nonumber
			&\!\!\!\!\!\!\Big(\mathcal{W}_c+\xi {\rm  Tr}\big(\mathbf{\Theta}\big(\mathbf{H} (\sum\nolimits_{k=1}^K\mathbf{f}_k\mathbf{f}_k^{\rm H})\mathbf{H}^{\rm H}+\sigma_1^2\mathbf{I}_M\big)\mathbf{\Theta}^{\rm H}\big)  \Big)\\
			\label{C_energy-1}
			&\!\!\!\!\!\!\times (1-\Omega)Z^{-1}+M\Omega  \leq \sum\nolimits_{m=1}^{M}\mathcal{C}_m^{-1},\\
			\label{C_energy-2}
			&\!\!\!\!\!\!\zeta_m \!\leq \!{\rm Tr}\Big(\! \mathbf{T}_m\mathbf{H}\big(\sum\nolimits_{k=1}^K\!\!\!\mathbf{f}_k\mathbf{f}_k^{\rm H}\big)\mathbf{H}^{\rm H}\mathbf{T}_m^{\rm H}\Big)\!\!+\!\!(1 \!-\!\alpha_{m})\sigma_1^2,~\forall m,\!\!\!\!\!	\\
			\label{C_energy-3}
			&\!\!\!\!\!\!\mathcal{C}_m\geq \exp\big(-a(\zeta_m-q)\big)+1,~\forall m,
		\end{align}
	\end{subequations}
	where $\mathcal{W}_c=\sum\nolimits_{m=1}^M\alpha_m(P_b+P_{\rm DC})+(M-\sum\nolimits_{m=1}^{M} \alpha_{m}) P_{\rm C}$.
    Since constraint (\ref{C_energy-1}) remains non-convex, we approximate it using the FTS. 
	For the feasible point $\{\mathcal{C}_m\}$ in the $\ell$-th iteration, a lower bound on $\sum\nolimits_{m=1}^{M}\mathcal{C}_m^{-1}$ is given by $\mathcal{C}_{\rm lb}=\sum\nolimits_{m=1}^M\big(2(\mathcal{C}_m^{(\ell)})^{-1}-\mathcal{C}_m(\mathcal{C}_m^{(\ell)})^{-2}\big)$.
	Now, defining $\Delta=\{Q_{k},\mathcal{A}_{k},\mathcal{B}_{k},\mathcal{C}_m,\zeta_m|\forall k,\forall m\}$ as an auxiliary variable set and denoting $\bar{\mathcal{W}}_c=\frac{(\mathcal{C}_{\rm lb}-M\Omega)Z}{(1-\Omega)\xi}-\frac{\mathcal{W}_c}{\xi}$, problem (\ref{P0}) is recast as
	\begin{subequations}
		\label{P0-1}
		\begin{eqnarray}
			&\!\!\!\!\!\!\!\!\!\!\!\!\!\!\!\!\!\underset{\mathbf{f}_k,\boldsymbol{\Theta},\Delta}{\max}  
			&\!\!\!\!\sum\nolimits_{k=1}^KQ_{k}\\
			\label{C_Q_lb}
			&\!\!\!\!\!\!\!\!\!\!\!\!\!\!\!\!\operatorname{s.t.} &\!\!\!\!Q_{k}\leq R_{k}^{\rm lb}, ~\forall k,\\
			\label{C_C_lb}
			&&\!\!\!\!\bar{\mathcal{W}}_c\geq {\rm  Tr}\big(\mathbf{\Theta}\big(\mathbf{H} (\sum\nolimits_{k=1}^K\mathbf{f}_k\mathbf{f}_k^{\rm H})\mathbf{H}^{\rm H}\!+\!\sigma_1^2\mathbf{I}_M\big)\mathbf{\Theta}^{\rm H}\big),\\
			&&\!\!\!\!{\rm (\ref{C_transmit beamforming}), (\ref{C-MF-RIS}),(\ref{C_AB_1}), (\ref{C_AB_2}), (\ref{C_energy-2}), (\ref{C_energy-3}). }
		\end{eqnarray}
	\end{subequations}

\begin{algorithm}[tbp]
\setstretch{1}
\caption{The SROCR-Based Algorithm for Solving (\ref{P_f-1})}
\label{Algorithm_active}
\begin{algorithmic}[1] 
	\STATE \textbf{Initialization:} set $\ell_1=0$, initialize the feasible point $\{\mathbf{F}_{k}^{(\ell_1)},w_{k}^{(\ell_1)}\}$ and the step size $\delta_1^{(\ell_1)}$.
	\REPEAT
	\STATE Solve the convex problem (\ref{P_f_2}) to obtain $\mathbf{F}_k$;
	\IF {problem (\ref{P_f_2}) is solvable} 
	\STATE 
	Update $\mathbf{F}_k^{(\ell+1)}=\mathbf{F}_k$ and
	$\delta_1^{(\ell_1+1)}=\delta_1^{(0)}$;
	\ELSE
	\STATE 
	Update $\mathbf{F}_k^{(\ell+1)}=\mathbf{F}_k^{(\ell)}$ and
	$\delta_1^{(\ell_1+1)}={\delta_1^{(\ell_1)}}/{2}$;
	\ENDIF
	\STATE Update $\ell_1=\ell_1+1$ and 
	\STATE Update $w_{k}^{(\ell_1)}=\min\Big(1,\frac{\lambda_{\rm max}(\mathbf{F}_{k}^{(\ell_1)})}
	{{\rm{Tr}}(\mathbf{F}_{k}^{(\ell_1)})}+\delta_1^{(\ell_1)}\Big)$;
	\UNTIL the stopping criterion is met.
\end{algorithmic}  
\end{algorithm}
	\subsection{Transmit Beamforming Under Perfect CSI}
	\vspace{-1mm}
	
	To solve problem (\ref{P0-1}), we define $\bar{\mathbf{H}}_{k}=\bar{\mathbf{h}}_{k}^{\rm H}\bar{\mathbf{h}}_{k}$ and $\mathbf{F}_k=\mathbf{f}_k\mathbf{f}_k^{\rm H}$, satisfying $\mathbf{F}_k\succeq \mathbf{0}$ and ${\rm rank}(\mathbf{F}_k)=1$.
	Then, constraints in (\ref{C_AB_1}), (\ref{C_AB_2}), (\ref{C_energy-2}), and (\ref{C_C_lb}) are, respectively, rewritten as
	\begin{subequations}
		\begin{align}
			\label{C_AB_f_1}
			&\!\!\!{\mathcal{A}_{k}^{-1}}\leq {\rm Tr}\big( \bar{\mathbf{H}}_{k}\mathbf{F}_k\big), ~\forall k,\\
			\label{C_AB_f_2}
			&\!\!\!\mathcal{B}_{k}\geq \sum\nolimits_{i=1,i\neq k}^K {\rm Tr}\big(\bar{\mathbf{H}}_{k}\mathbf{F}_{i}\big)+\sigma_1^2\lVert\mathbf{g}_{k}^{\mathrm H}\boldsymbol{\Theta}\lVert^2+\sigma_0^2,~\forall k,\\
			\label{C_f-energy-1}
			&\!\!\!\zeta_m \!\leq \!{\rm Tr}\big( \mathbf{T}_m\mathbf{H}(\sum\nolimits_{k=1}^K\!\!\mathbf{F}_k)\mathbf{H}^{\rm H}\mathbf{T}_m^{\rm H}\big)\! + \!(1\!-\!\alpha_{m})\sigma_1^2,~\forall m,\!\!\!\\
			\label{C_f-energy-2}
			& \!\!\!\bar{\mathcal{W}}_c\geq {\rm  Tr}\big(\mathbf{\Theta}\big(\mathbf{H} (\sum\nolimits_{k=1}^K\mathbf{F}_k)\mathbf{H}^{\rm H}+\sigma_1^2\mathbf{I}_M\big)\mathbf{\Theta}^{\rm H}\big).
		\end{align}
 	\end{subequations}
 
	Accordingly, with fixed $\boldsymbol{\Theta}$, the transmit beamforming subproblem is given by
	\setlength{\abovedisplayskip}{1pt}
	\setlength{\belowdisplayskip}{1pt}
	\begin{subequations}
		\label{P_f-1}
		\begin{eqnarray}
			&\underset{\mathbf{F}_k,\Delta}{\max}  &\sum\nolimits_{k=1}^KQ_{k}\\
			\label{C_f_Tr_0}
			&\operatorname{s.t.} &\sum\nolimits_{k=1}^K {\rm Tr}\big(\mathbf{F}_k\big) \leq P_{\rm BS}^{\max}, \\
			\label{C_f_rank}
			&&{\rm rank}(\mathbf{F}_k)=1, ~\forall k,\\
			\label{C_f_rank_0}
			&&\mathbf{F}_k\succeq \mathbf{0}, ~\forall k, ~{\rm (\ref{C_energy-3}),(\ref{C_Q_lb}), (\ref{C_AB_f_1}){\text -}(\ref{C_f-energy-2})}.
		\end{eqnarray}
	\end{subequations}

	To handle the rank-one constraint (\ref{C_f_rank}), we adopt the sequential rank-one constraint relaxation (SROCR) method\cite{SROCR-confer}.
	Specifically, by denoting $w_{k}^{(\ell-1)}\in[0,1]$ as the trace ratio parameter in the $(\ell-1)$-th iteration, constraint (\ref{C_f_rank}) is replaced by the following linear constraint:
	\setlength{\abovedisplayskip}{2pt}
	\setlength{\belowdisplayskip}{2pt}
	\begin{eqnarray}
		\label{SROCR-linear-f}
		\big(\mathbf{f}_{k}^{{\rm eig},(\ell-1)}\big)^{\rm H}\mathbf{F}_{k}^{(\ell)}\mathbf{f}_{k}^{{\rm eig},(\ell-1)}\geq w_{k}^{(\ell-1)}{\rm Tr}\big(\mathbf{F}_{k}^{(\ell)}\big), ~\forall k, 
	\end{eqnarray}
	where $\mathbf{f}_{k}^{{\rm eig},(\ell-1)}$ is the eigenvector corresponding to the largest eigenvalue of $\mathbf{F}_{k}^{(\ell-1)}$, and $\mathbf{F}_{k}^{(\ell-1)}$ is the obtained solution in the $(\ell\!-\!1)$-th iteration. 
	Thus, problem (\ref{P_f-1}) is rewritten as
	\setlength{\abovedisplayskip}{1pt}
	\setlength{\belowdisplayskip}{1pt}
	\begin{subequations}
		\label{P_f_2}
		\begin{eqnarray}
			&\underset{\mathbf{F}_k,\Delta}{\max}  &\sum\nolimits_{k=1}^KQ_{k}\\
			&\operatorname{s.t.} &{\rm (\ref{C_f_Tr_0}), (\ref{C_f_rank_0}), (\ref{SROCR-linear-f}). }
		\end{eqnarray}
	\end{subequations}
	Since problem (\ref{P_f_2}) is a standard semi-definite programming (SDP) problem, it can be solved efficiently via standard convex solver software, such as CVX\cite{CVX}. 
	The details	of solving (\ref{P_f-1}) are given in Algorithm \ref{Algorithm_active}.
	Specifically, $w_{k}^{(\ell-1)}=0$ indicates that the rank-one constraint is dropped, while $w_{k}^{(\ell-1)}=1$ is	equivalent to the rank-one constraint.
	Therefore, by increasing $w_{k}^{(\ell-1)}$ from $0$ to $1$ after each iteration, we can gradually approach a rank-one solution\cite{SROCR-confer}.
	After solving (\ref{P_f-1}), the solution of $\mathbf{f}_{k}$ can be obtained by using the Cholesky decomposition, i.e., $\mathbf{F}_{k}=\mathbf{f}_{k}\mathbf{f}_{k}^{\rm H}$.

	\subsection{MF-RIS Coefficient Deign Under Perfect CSI}\label{MF-RIS Coefficient Design}
	Next, we focus on the design of MF-RIS coefficients with given $\mathbf{f}_k$.
	First, we denote $\widetilde{\mathbf{H}}_{k}=\big[{\rm diag}(\mathbf{g}_{k}^{\rm H})\mathbf{H}; \mathbf{h}_{k}^{\rm H}\big]$ and $\mathbf{u}=\Big[\big[\alpha_1\sqrt{\beta_1}e^{j\theta_1},\alpha_2\sqrt{\beta_2}e^{j\theta_2},\cdots,\alpha_M\sqrt{\beta_M}e^{j\theta_M}\big]^{\rm H};1\Big]$.
	We further define $\mathbf{U}\!=\!\mathbf{u}\mathbf{u}^{\rm H}$, satisfying $\mathbf{U}\succeq \mathbf{0}$, ${\rm rank}(\mathbf{U})\!=\!1$,  $\left[\mathbf{U}\right]_{m, m}\!=\!\alpha_m^2\beta_{m}$, and $\left[\mathbf{U}\right]_{M+1, M+1}\!=\!1$.
	Then, we have
	\setlength{\abovedisplayskip}{3pt}
	\setlength{\belowdisplayskip}{3pt}
	\begin{align}
		\label{C_A_H_passive}
		|\bar{\mathbf{h}}_{k}\mathbf{f}_{k}|^2=|(\mathbf{h}_{k}^{\rm H}+\mathbf{g}_{k}^{\rm H}\boldsymbol{\Theta}\mathbf{H})\mathbf{f}_{k}|^2=\operatorname{Tr}(\widetilde{\mathbf{H}}_{k}\mathbf{F}_k\widetilde{\mathbf{H}}_{k}^{\rm H}\mathbf{U}).
	\end{align}
	
	Similarly, by defining $\bar{\mathbf{G}}_{k}=\bar{\mathbf{g}}_{k}\bar{\mathbf{g}}_{k}^{\rm H}$ and $\bar{\mathbf{H}}=\bar{\mathbf{h}}\bar{\mathbf{h}}^{\rm H}+\sigma_1^2\bar{\mathbf{I}}_M\bar{\mathbf{I}}_M^{\rm H}$, with
	$\bar{\mathbf{g}}_{k}=[\sigma_1{\rm diag}(\mathbf{g}_{k}^{\rm H}); \mathbf{0}_{1\times M}]$, $\bar{\mathbf{h}}=[\mathbf{H}\mathbf{w}; 0]$, and $\bar{\mathbf{I}}_{M}=[\mathbf{I}_M;\mathbf{0}_{1\times M}]$, the term $\sigma_1^2\lVert\mathbf{g}_{k}^{\mathrm H}\boldsymbol{\Theta}\lVert^2$ and the output power of MF-RIS are rewritten as $\sigma_1^2\lVert\mathbf{g}_{k}^{\mathrm H}\boldsymbol{\Theta}\lVert^2=\operatorname{Tr}(\bar{\mathbf{G}}_{k}\mathbf{U})$ and
	$P_{\rm O}=\operatorname{Tr}( \bar{\mathbf{H}}\mathbf{U})$, respectively.
	Constraints (\ref{C_AB_1}), (\ref{C_AB_2}), and (\ref{C_C_lb}) are then, respectively, transformed into 
	\begin{subequations}
		\begin{align}
			\label{C_AB_passive_1}
			&\mathcal{A}_{k}^{-1} \geq \operatorname{Tr}(\widetilde{\mathbf{H}}_{k}\mathbf{F}_k\widetilde{\mathbf{H}}_{k}^{\rm H}\mathbf{U}), \\
			\label{C_AB_passive_2}
			&\mathcal{B}_{k}\geq \sum\nolimits_{i=1,i\neq k}^K  \operatorname{Tr}(\widetilde{\mathbf{H}}_{k}\mathbf{F}_{i}\widetilde{\mathbf{H}}_{k}^{\rm H}\mathbf{U})+\operatorname{Tr}(\bar{\mathbf{G}}_{k}\mathbf{U})+\sigma_0^2, \\
			\label{C_AB_passive_3}
			&\bar{\mathcal{W}}_c\geq \operatorname{Tr}( \bar{\mathbf{H}}\mathbf{U}). 
		\end{align}
	\end{subequations}

   Consequently, the MF-RIS coefficient design problem is reformulated as
	\begin{subequations}
		\label{P_passive-1}
		\begin{eqnarray}
			\label{P_passive-1-objective function}
			&\underset{\mathbf{U},\Delta}{\max}  &\sum\nolimits_{k}Q_{k}\\
			&\operatorname{s.t.} 
			\label{C_passive_rank_2}
			&\left[\mathbf{U}\right]_{m, m}=\alpha_m^2\beta_{m}, ~\forall m, \\
			\label{C_passive_rank_3}
			&&\mathbf{U}\succeq \mathbf{0}, ~\left[\mathbf{U}\right]_{M+1, M+1}=1, \\
			\label{C_passive_rank_4}
			&&{\rm rank}(\mathbf{U})=1, \\
			\label{C_passive-alpha}
			&&\alpha_{m}\in\{0,1\}, ~\forall m,\\
			\label{C-passive-beta-C-1}
			&& \beta_{m}\in\left[0,\beta_{\max}\right], ~\forall m, \\
			\label{C-passive-beta-C-2}
			&&{\rm (\ref{C_energy-2}),(\ref{C_energy-3}), (\ref{C_Q_lb}),(\ref{C_AB_passive_1}){\text -}(\ref{C_AB_passive_3})}. 
		\end{eqnarray}
	\end{subequations}

	Problem (\ref{P_passive-1}) is intractable due to the highly-coupled constraint (\ref{C_passive_rank_2}), the rank-one constraint (\ref{C_passive_rank_4}), and the binary constraint (\ref{C_passive-alpha}).
	Similar to the transformation of the rank-one constraint (\ref{C_f_rank}), we adopt the SROCR method to tackle (\ref{C_passive_rank_4}).
    Specifically, we denote $v^{(\ell-1)}\in[0,1]$ as the trace ratio parameter of $\mathbf{U}$ in the $(\ell-1)$-th iteration, $\mathbf{u}^{{\rm eig},(\ell-1)}$ as the eigenvector corresponding to the largest eigenvalue of $\mathbf{U}^{(\ell-1)}$, and $\mathbf{U}^{(\ell-1)}$ as the obtained solution in the $(\ell-1)$-th iteration with $v^{(\ell-1)}$.
	Constraint (\ref{C_passive_rank_4}) in the $\ell$-th iteration then becomes the following linear one:
	\begin{align}
	\label{SROCR-linear-theta}
	\big(\mathbf{u}^{{\rm eig},(\ell-1)}\big)^{\rm H}\mathbf{U}^{(\ell)}\mathbf{u}^{{\rm eig},(\ell-1)}
	\geq v^{(\ell-1)}{\rm Tr}(\mathbf{U}^{(\ell)}).
   \end{align}

	As for the binary constraint (\ref{C_passive-alpha}),  we equivalently recast it into two continuous ones: 
	 $\alpha_m- \alpha_m^2\leq 0$ and $0 \leq \alpha_m  \leq 1$.
    The SCA technique is employed to address the non-convex constraint $\alpha_m-\alpha_m^2  \leq  0$.
	For the feasible point $\{\alpha_m^{(\ell)}\}$ in the $\ell$-th iteration, using the FTS, a convex upper bound on $-\alpha_m^2$ is obtained as 
	$\left(-\alpha_m^2\right)^{\rm ub} = -2\alpha_m^{(\ell)}\alpha_m  + (\alpha_m^{(\ell)})^2$.
	\begin{algorithm}[tbp]
	\setstretch{1}
	\caption{The Penalty Function-Based Algorithm for Solving Problem (\ref{P_passive-1})}
	\label{Algorithm_passive}
	\begin{algorithmic}[1] 
		\STATE \textbf{Initialization:} set the	initial iteration index $\ell_2=0$, 
		initialize the feasible point $\{\mathbf{U}^{(\ell_2)},v_{k}^{(\ell_2)}\}$, $\varepsilon>1$, and the step size $\delta_2^{(\ell_2)}$.
		\REPEAT
		\IF {$\ell_2\leq T_{\max} $} 
		\STATE Solve the convex problem (\ref{P_passive-2}) to obtain $\mathbf{U}$;
		\IF {problem (\ref{P_passive-2}) is solvable} 
		\STATE Update $\mathbf{U}^{(\ell+1)}=\mathbf{U}$ and $\delta_2^{(\ell_2+1)}=\delta_2^{(0)}$;
		\ELSE
		\STATE Update $\mathbf{U}^{(\ell+1)}=\mathbf{U}^{(\ell)}$ and $\delta_2^{(\ell_2+1)}={\delta_2^{(\ell_2)}}/{2}$.
		\ENDIF
		\STATE Update $\ell_2=\ell_2+1$;
		\STATE Update $v_{k}^{(\ell_2)}=\min\Big(1,\frac{\lambda_{\rm max}(\mathbf{U}^{(\ell_2)})}
		{{\rm{Tr}}(\mathbf{U}^{(\ell_2)})}+\delta_2^{(\ell_2)}\Big)$;
		\STATE
		Update $\rho^{(\ell_2)}={\min}\{\varepsilon\rho^{(\ell_2-1)}, \rho_{\rm max}\}$;
		\ELSE 
		\STATE Reinitialize with a new feasible point $\{\mathbf{U}^{(0)},v_{k}^{(0)}\}$, set $\varepsilon>1$ and $\ell_2=0$.
		\ENDIF
		\UNTIL the stopping criterion is met.
	\end{algorithmic}  
\end{algorithm}
	
	By introducing the auxiliary variable, $\eta_m=\alpha_{m}^2\beta_{m}$, the highly-coupled non-convex constraint (\ref{C_passive_rank_2}) is recast as
	\begin{align}
		\label{C_passive_rank_2-relax}
		\left[\mathbf{U}\right]_{m, m}=\eta_m, ~~\eta_m\leq \alpha_{m}^2\beta_{m}, ~~\alpha_{m}^2\beta_{m} \leq \eta_m, ~\forall m.
	\end{align}
	Next, we apply the penalty function-based method to deal with the non-convex constraints $\eta_m\leq \alpha_{m}^2\beta_{m}$ and $\alpha_{m}^2\beta_{m} \leq \eta_m$.
	The former is approximated by $\eta_m\leq 2(\alpha_{m}-\alpha_{m}^{(\ell)})\alpha_{m}^{(\ell)}\beta_{m}^{(\ell)}+(\alpha_{m}^{(\ell)})^2\beta_{m}$, where the RHS is the FTS of $\alpha_{m}^2\beta_{m}$ at the feasible point $\{\alpha_{m}^{(\ell)},\beta_{m}^{(\ell)}\}$ obtained in the $\ell$-th iteration.
	As for the latter, we replace the term $\alpha_{m}^2\beta_{m}$ with its convex upper bound.
	Specifically, defining the functions $g(\alpha_{m},\beta_{m})=\alpha_{m}^2\beta_{m}$ and $G(\alpha_{m},\beta_{m})=\frac{c_m}{2}\alpha_{m}^4+\frac{\beta_{m}^2}{2c_m}$, $G(\alpha_{m},\beta_{m})\geq g(\alpha_{m},\beta_{m})$ is then satisfied for $\alpha_{m},\beta_{m},c_m>0$.  
	When $c_m=\frac{\beta_{m}}{\alpha_{m}^2}$, the equations $g(\alpha_{m},\beta_{m})=G(\alpha_{m},\beta_{m})$ and $\nabla g(\alpha_{m},\beta_{m})=\nabla G(\alpha_{m},\beta_{m})$ hold\cite{Throughput_NOMA_TWC}.
	Eventually, problem (\ref{P_passive-1}) is reformulated as
	\begin{subequations}
		\label{P_passive-2}
		\begin{align}
			\underset{\mathbf{U},\Delta,\boldsymbol{\eta},\mathbf{d}}{\max}  ~&\sum\nolimits_{k=1}^KQ_{k}-\rho\sum\nolimits_{m=1}^M (d_m+\bar{d}_m) \\
			\operatorname{s.t.} ~&\left[\mathbf{U}\right]_{m, m}=\eta_m, ~\forall m,\\
			\label{C-passive-alpha}
			&0\leq \alpha_m \leq 1, ~\alpha_m+\left(-\alpha_m^2\right)^{\rm ub}\leq 0, ~\forall m,\\	
			\nonumber
			& \eta_m\leq 2(\alpha_{m}-\alpha_{m}^{(\ell)})\alpha_{m}^{(\ell)}\beta_{m}^{(\ell)}\\
			&~~~~~~~+(\alpha_{m}^{(\ell)})^2\beta_{m}+d_m, ~\forall m,\\
			\label{C-passive-alpha-beta}
			&\frac{c_m}{2}\alpha_{m}^4+\frac{\beta_{m}^2}{2c_m}\leq \eta_m+\bar{d}_m, ~\forall m,\\
			&{\rm (\ref{C_passive_rank_3}),(\ref{C-passive-beta-C-1}), (\ref{C-passive-beta-C-2}),(\ref{SROCR-linear-theta})},
		\end{align}
	\end{subequations}
	where $\boldsymbol{\eta}=\{\eta_m|\forall m\}$. 
	The set $\mathbf{d}=\{d_m,\bar{d}_m|\forall m\}$ is a slack variable set imposed over the non-convex constraints $\eta_m\leq \alpha_{m}^2\beta_{m}$ and $\alpha_{m}^2\beta_{m} \leq \eta_m$, and $\rho$ is a penalty factor used to penalize the violation of these constraints.
	Problem (\ref{P_passive-2}) is a convex SDP, and thus can be solved efficiently via CVX\cite{CVX}.
	The fixed point $c_m$ in the $\ell$-th iteration is updated by $c_m^{(\ell)}=\frac{\beta_{m}^{(\ell-1)}}{(\alpha_{m}^{(\ell-1)})^2}$.
	The details of the proposed penalty function-based algorithm are given in Algorithm \ref{Algorithm_passive}.
	
	\vspace{-1mm}
	\subsection{Complexity and Convergence Analysis}
    Based on the AO framework, the solution of problem (\ref{P0}) can be obtained by solving problem (\ref{P_f-1}) and problem (\ref{P_passive-1}) alternately.
	The complexity for solving problem (\ref{P_f-1}) and problem (\ref{P_passive-1}) with the interior-point method is given by $\mathcal{O}_{\mathbf{f}}=\mathcal{O}\big(I_{\rm ite}^{\mathbf{f}}\max (KN, 2K+M)^4\sqrt{KN}\big)$ and $\mathcal{O}_{\boldsymbol{\Theta}}=\mathcal{O}\big(I_{\rm ite}^{\boldsymbol{\Theta}} $  $\max (M+1, 2K+M)^4\sqrt{M+1}\big)$, respectively, 
	where $I_{\rm ite}^{\mathbf{f}}$ and $I_{\rm ite}^{\boldsymbol{\Theta}}$ denote the corresponding numbers of iterations\cite{Throughput_NOMA_MU-TWC}. 
   The convergence of the overall algorithm is analyzed as follows.
	
   	Define $U\big(\mathbf{f}_k^{(\ell)},\boldsymbol{\Theta}^{(\ell)}\big)$ as the objective function value of problem (\ref{P0}) in the $\ell$-th iteration.
   Then, for the transmit beamforming optimization problem (\ref{P_f_2}) with a given $\boldsymbol{\Theta}^{(\ell)}$, we have the following inequalities
   \begin{align}
   	\nonumber
   	U\big(\mathbf{f}_k^{(\ell)},\boldsymbol{\Theta}^{(\ell)}\big)
   	&\overset{(a)}{=} U^{\rm lb}_{\mathbf{f}_k^{(\ell)}}\big( \mathbf{f}_k^{(\ell)},\boldsymbol{\Theta}^{(\ell)}\big)
   	\overset{(b)}{\leq} U^{\rm lb}_{\mathbf{f}_k^{(\ell)}}\big( \mathbf{f}_k^{(\ell+1)},\boldsymbol{\Theta}^{(\ell)}\big) \\
   	\label{convergence_1}
   	&\overset{(c)}{\leq} U\big(\mathbf{f}_k^{(\ell+1)},\boldsymbol{\Theta}^{(\ell)}\big), 
   \end{align}
   where $U_{\mathbf{f}_k^{(\ell)}}^{\rm lb}$ denotes the objective function value of problem (\ref{P_f_2}) for the local point $\{\mathbf{f}_k^{(\ell)}\}$.
   Here, $(a)$ holds because the FTS is tight at the given local point\cite{Throughput_NOMA_MU-TWC}, $(b)$ follows from the fact that the solution of $\{\mathbf{f}_k^{(\ell+1)}\}$ is obtained via Algorithm \ref{Algorithm_active} with a given $\boldsymbol{\Theta}^{(\ell)}$, and $(c)$ is because problem (\ref{P_f_2}) always provides a lower-bound solution for the original problem (\ref{P0}).
   Therefore, for fixed $\boldsymbol{\Theta}^{(\ell)}$, the objective function value of problem (\ref{P0}) is non-decreasing after solving problem (\ref{P_f_2}). 
   Similarly, for the MF-RIS coefficient design problem (\ref{P_passive-2}), we have
   \begin{align}
   	\nonumber
   	U\big(\mathbf{f}_k^{(\ell+1)},\boldsymbol{\Theta}^{(\ell)}\big)
   	&\! \!=\! U^{\rm lb}_{\boldsymbol{\Theta}^{(\ell)}}\big( \mathbf{f}_k^{(\ell+1)},\boldsymbol{\Theta}^{(\ell)}\big)
   	\!\!   \leq \!  U^{\rm lb}_{\boldsymbol{\Theta}^{(\ell)}}\big( \mathbf{f}_k^{(\ell+1)},\boldsymbol{\Theta}^{(\ell+1)}\big) \\
   	\label{convergence_2}
   	&\!  \leq \!  U\big(\mathbf{f}_k^{(\ell+1)},\boldsymbol{\Theta}^{(\ell+1)}\big),
   \end{align}
   where $U^{\rm lb}_{\boldsymbol{\Theta}^{(\ell)}}$ is the objective function value of problem (\ref{P_passive-2}) at a local point $\{\boldsymbol{\Theta}^{(\ell)}\}$.
   Based on (\ref{convergence_1}) and (\ref{convergence_2}), it holds that
   \begin{align}
   	\label{convergence}
   	U\big(\mathbf{f}_k^{(\ell)},\boldsymbol{\Theta}^{(\ell)}\big)
   	\leq U\big(\mathbf{f}_k^{(\ell+1)},\boldsymbol{\Theta}^{(\ell+1)}\big).  
   \end{align} 
   Inequality (\ref{convergence}) indicates that the objective function value of problem (\ref{P0}) is monotonically non-decreasing after each iteration.
   Besides, constraints (\ref{C_transmit beamforming}) and (\ref{C-MF-RIS}) limit the maximum transmit power at the BS and the maximum amplification factor at the MF-RIS, respectively. 
   The limited number of RIS elements and the logistic function-based energy harvesting model (\ref{Energy_harvesting_model}) restrict the maximum available power at the MF-RIS, thus ensuring the boundness of the objective function. 
	Hence, the AO algorithm is guaranteed to converge.

	\section{Throughput Maximization Under Imperfect CSI}\label{Imperfect CSI case}
	The acquisition of perfect CSI is challenging due to inevitable channel estimation and quantization errors.
	Therefore, in this section, we propose a robust beamforming scheme by taking into account the imperfect CSI.
   
   \vspace{-2mm}
	\subsection{Problem Formulation Under Imperfect CSI}
	Considering that the acquired CSI is coarse and outdated, we adopt the bounded CSI model to characterize the uncertainties of CSI, given by\cite{Secure_Wang_JSTSP}
	\begin{subequations}
		\begin{align}
			&\mathbf{h}_k=\widetilde{\mathbf{h}}_k+ \triangle \mathbf{h}_k, 
			~~\mathbf{g}_k=\widetilde{\mathbf{g}}_k+  \triangle \mathbf{g}_k,  ~\forall k,\\
			&\mathbf{H}=\widetilde{\mathbf{H}}+  \triangle \mathbf{H}, 
			~~\mathbf{G}_k=\widetilde{\mathbf{G}}_k+  \triangle \mathbf{G}_k, ~\forall k, \\
			&\Lambda_{h,k}=\{  \triangle \mathbf{h}_k\in\mathbb{C}^{N\times 1}: \lVert  \triangle \mathbf{h}_k \lVert \leq \xi_{h,k}\}, ~\forall k,\\
			&\Lambda_{g,k}=\{  \triangle \mathbf{g}_k\in\mathbb{C}^{M\times 1}: \lVert  \triangle \mathbf{g}_k \lVert \leq \xi_{g,k}\}, ~\forall k,\\
			&\Lambda_{H}=\{  \triangle \mathbf{H}\in\mathbb{C}^{M\times N}: ~ \lVert  \triangle \mathbf{H} \lVert_F\leq \xi_{H}\}, ~\forall k,\\
			&\Lambda_{G,k}=\{ \triangle \mathbf{G}_k\in\mathbb{C}^{M\times N}: \lVert  \triangle \mathbf{G}_k \lVert_F \leq \xi_{G,k} \}, ~\forall k,
		\end{align}
	\end{subequations}
	where $\mathbf{G}_k={\rm diag}(\mathbf{g}_k^{\rm H})\mathbf{H}$ is the cascaded channel from the BS to user $k$.
	Here, $\widetilde{\mathbf{h}}_k$ and $\triangle \mathbf{h}_k$ are the estimate of channel $\mathbf{h}_k$ and the corresponding estimation error, respectively.
    The continuous set $\Lambda_{h,k}$ collects all possible estimation errors, with $\xi_{h,k}>0$ denoting the radii of the uncertainty regions.
	Other parameters, $\widetilde{\mathbf{g}}_k$, $\widetilde{\mathbf{H}}$,  $\widetilde{\mathbf{G}}_k$, $ \triangle \mathbf{g}_k$, $\triangle \mathbf{H}$, and $\triangle\mathbf{G}_{k}$, are defined similarly.
    Similar to problem (\ref{P0-1}), the SR maximization problem in the imperfect CSI case is formulated as
	\begin{subequations}
		\label{P0-robust-1}
		\begin{eqnarray}
			&\!\!\!\!\!\!\!\!\!\!\!\!\!\!\!\!\!\!\!\underset{\mathbf{f}_k,\boldsymbol{\Theta},\Delta}{\max}  &\!\!\!\!\!\!\sum\nolimits_{k=1}^KQ_{k}\\
			\label{C_A_robust}
			&\!\!\!\!\!\!\!\!\!\!\!\!\!\!\!\!\!\!\!\operatorname{s.t.} &\!\!\!\!\!\!\!{\mathcal{A}_{k}^{-1}}\leq |\bar{\mathbf{h}}_{k}\mathbf{f}_{k}|^2,  ~\Lambda_{h,k}, \Lambda_{G,k},\forall k,\\
			\nonumber
			&&\!\!\!\!\!\!\!\mathcal{B}_{k}\geq \sum\nolimits_{i=1,i\neq k}^K|\bar{\mathbf{h}}_{k}\mathbf{f}_{i}|^2+\sigma_1^2\lVert\mathbf{g}_{k}^{\mathrm H}\boldsymbol{\Theta}\lVert^2+\sigma_0^2, \\
			\label{C_B_robust}
			&&\!\!\!\!\!\!\!\Lambda_{h,k}, \Lambda_{g,k},\Lambda_{G,k},  \forall k,\\
			\nonumber
			&&\!\!\!\!\!\!\!\zeta_m \leq {\rm Tr}\Big( \mathbf{T}_m\mathbf{H}\big(\sum\nolimits_{k=1}^K\mathbf{f}_k\mathbf{f}_k^{\rm H}\big)\mathbf{H}^{\rm H}\mathbf{T}_m^{\rm H}\Big)\\
			\label{C_P_RF_robust}
			&&\!\!\!\!\!\!\!~~~~~~~+(1-\alpha_{m})\sigma_1^2, ~\Lambda_{H},\forall m,	\\
			\label{C_P_O_robust}
			&&\!\!\!\!\!\!\!\bar{\mathcal{W}}_c\!\geq \!{\rm  Tr}\Big(\!\mathbf{\Theta}\big(\mathbf{H}(\sum\nolimits_{k=1}^K\!\!\mathbf{f}_k\mathbf{f}_k^{\rm H})\mathbf{H}^{\rm H} \!\!\!+\!\!\sigma_1^2\mathbf{I}_M\big)\mathbf{\Theta}^{\rm H}\!\Big), \!\Lambda_{H}, \\
			&&\!\!\!\!\!\!\!{\rm (\ref{C_transmit beamforming}), (\ref{C-MF-RIS}), (\ref{C_energy-3}), (\ref{C_Q_lb})}.
		\end{eqnarray}
	\end{subequations}

	Compared to problem (\ref{P0-1}), the difficulty of solving problem (\ref{P0-robust-1}) lies in the infinitely many non-convex constraints (\ref{C_A_robust})-(\ref{C_P_O_robust}) caused by the CSI imperfectness.
	To this end, we use the $\mathcal{S}$-procedure and the general sign-definiteness to transform (\ref{C_A_robust})-(\ref{C_P_O_robust}) into tractable forms.
	The AO framework is then utilized to decompose the reformulated problem into two subproblems, and we further optimize the transmit beamforming and MF-RIS coefficients alternately.
	
	\vspace{-2mm}
	\subsection{Problem Transformation Under Imperfect CSI}
	To deal with constraint (\ref{C_A_robust}), we first derive its linear approximation in the following lemma.

	\begin{lemma}
		\label{Lemma-SCA}
		\emph{
			By denoting $(\mathbf{f}_k^{(\ell)}, \mathbf{v}^{(\ell)})$ as the solution obtained in the $\ell$-th iteration and defining $\mathbf{v}=$ $\big[\alpha_1\sqrt{\beta_1}e^{j\theta_1},\alpha_2\sqrt{\beta_2}e^{j\theta_2},\cdots,\alpha_M\sqrt{\beta_M}e^{j\theta_M}\big]^{\rm T}$, constraint (\ref{C_A_robust}) is equivalently linearized by
			\begin{eqnarray}	   
				\label{robust_A_SCA}
				\mathbf{x}_{k}^{\mathrm H}\mathbf{A}_{k}\mathbf{x}_{k}+2{\rm Re}\{\mathbf{a}_{k}^{\mathrm H}\mathbf{x}_{k}\}+a_{k} \geq {\mathcal{A}_{k}^{-1}}, ~\Lambda_{h,k},\Lambda_{G,k}, ~\forall k, 
			\end{eqnarray}
			where the vector $\mathbf{x}_{k}$ and the introduced coefficients $\mathbf{A}_{k}$, $\mathbf{a}_{k}$, and $a_{k}$ are given by (\ref{robust_A_SCA_coefficients}) at the top of the next page.	
		}
	\end{lemma}
	\begin{IEEEproof}
		Please refer to Appendix \ref{proof_of_Lemma_SCA}.
	\end{IEEEproof}
			\begin{figure*}[t]
	\begin{subequations}
		\label{robust_A_SCA_coefficients}
		\begin{eqnarray}
			\label{rho}
			\!\!\!\!\!\!\!\!\!\!\!\!\!\!\!\mathbf{x}_{k}\!\!\!\!\!\!\!\!\!\!&{}&=\Big[ \triangle\mathbf{h}_{k}^{\mathrm H}\ {\rm vec}^{\mathrm H}(\triangle \mathbf{G}_{k}^{\ast})\Big]^{\mathrm H},
			~~\mathbf{A}_{k}=\mathbf{\widetilde{A}}_{k}+\mathbf{\widetilde{A}}_{k}^{\mathrm H}-\mathbf{\widehat{A}}_{k},
			~~\mathbf{a}_{k}=\mathbf{\widetilde{a}}_{k}+\mathbf{\widehat{a}}_{k}-\mathbf{\bar{a}}_{k},
			~~a_{k}=2{\rm Re}\{\widetilde{a}_{k}\}-\widehat{a}_{k},\\
			\!\!\!\!\!\!\!\!\!\!\!\!\mathbf{\widetilde{A}}_{k}\!\!\!\!\!\!\!\!\!\!&{}&=\left[\begin{array}{c}
				\mathbf{f}_{k}^{(\ell)} \\
				\mathbf{f}_{k}^{(\ell)} \otimes (\mathbf{v}^{(\ell)})^{\ast}
			\end{array}\right]\Big[\mathbf{f}_{k}^{\mathrm H} \  \mathbf{f}_{k}^{\mathrm H} \otimes \mathbf{v}^{\mathrm T}\Big],
			~~\mathbf{\widehat{A}}_{k}=\left[\begin{array}{c}
				\mathbf{f}_{k}^{(\ell)} \\
				\mathbf{f}_{k}^{(\ell)} \otimes (\mathbf{v}^{(\ell)})^{\ast}
			\end{array}\right]\Big[(\mathbf{f}_{k}^{(\ell)})^{\mathrm H} \  (\mathbf{f}_{k}^{(\ell)})^{\mathrm H} \otimes (\mathbf{v}^{(\ell)})^{\mathrm T}\Big],\\
			\!\!\!\!\!\!\!\!\!\!\!\!\mathbf{\widetilde{a}}_{k}\!\!\!\!\!\!\!\!\!\!&{}&=\left[\begin{array}{c}
				\mathbf{f}_{k}(\mathbf{f}_{k}^{(\ell)})^{\mathrm H}(\widetilde{\mathbf{h}}_{k}+\widetilde{\mathbf{G}}_{k}^{\mathrm H}\mathbf{v}^{(\ell)}) ;
				{\rm vec}^{\ast}(\mathbf{v}(\widetilde{\mathbf{h}}_{k}^{\mathrm H}+(\mathbf{v}^{(\ell)})^{\mathrm H}\widetilde{\mathbf{G}}_{k})\mathbf{f}_{k}^{(\ell)}\mathbf{f}_{k}^{\mathrm H}
				)\end{array}\right], \\
			\!\!\!\!\!\!\!\!\!\!\!\!\mathbf{\widehat{a}}_{k}\!\!\!\!\!\!\!\!\!\!&{}&=\left[\begin{array}{c}
				\mathbf{f}_{k}^{(\ell)}\mathbf{f}_{k}^{\mathrm H}(\widetilde{\mathbf{h}}_{k}+\widetilde{\mathbf{G}}_{k}^{\mathrm H}\mathbf{v}) ; 
				{\rm vec}^{\ast}(\mathbf{v}^{(\ell)}(\widetilde{\mathbf{h}}_{k}^{\mathrm H}+\mathbf{v}^{\mathrm H}\widetilde{\mathbf{G}}_{k})\mathbf{f}_{k}(\mathbf{f}_{k}^{(\ell)})^{\mathrm H}
				)\\
			\end{array}\right],\\
			\!\!\!\!\!\!\!\!\!\!\!\!\mathbf{\bar{a}}_{k}\!\!\!\!\!\!\!\!\!\!&{}&=\left[\begin{array}{c}
				\mathbf{f}_{k}^{(\ell)}(\mathbf{f}_{k}^{(\ell)})^{\mathrm H}(\widetilde{\mathbf{h}}_{k}+\widetilde{\mathbf{G}}_{k}^{\mathrm H}\mathbf{v}^{(\ell)}) ;
				{\rm vec}^{\ast}(\mathbf{v}^{(\ell)}(\widetilde{\mathbf{h}}_{k}^{\mathrm H}+(\mathbf{v}^{(\ell)})^{\mathrm H}\widetilde{\mathbf{G}}_{k})\mathbf{f}_{k}^{(\ell)}(\mathbf{f}_{k}^{(\ell)})^{\mathrm H}
				)\\
			\end{array}\right],\\
			\!\!\!\!\!\!\!\!\!\widetilde{a}_{k}\!\!\!\!\!\!\!\!\!\!&{}&=(\widetilde{\mathbf{h}}_{k}^{\mathrm H}+(\mathbf{v}^{(\ell)})^{\mathrm H}\widetilde{\mathbf{G}}_{k})\mathbf{f}_{k}^{(\ell)}\mathbf{f}_{k}^{\mathrm H}(\widetilde{\mathbf{h}}_{k}+\widetilde{\mathbf{G}}_{k}^{\mathrm H}\mathbf{v}),
			~~\widehat{a}_{k}=(\widetilde{\mathbf{h}}_{k}^{\mathrm H}+(\mathbf{v}^{(\ell)})^{\mathrm H}\widetilde{\mathbf{G}}_{k})\mathbf{f}_{k}^{(\ell)}(\mathbf{f}_{k}^{(\ell)})^{\mathrm H}(\widetilde{\mathbf{h}}_{k}+\widetilde{\mathbf{G}}_{k}^{\mathrm H}\mathbf{v}^{(\ell)}).
		\end{eqnarray}
	\end{subequations}
	\hrulefill
\end{figure*}

    The linear constraint (\ref{robust_A_SCA}) still has infinite possibilities. To facilitate the derivation, we resort to the $\mathcal{S}$-procedure to further convert it into a manageable form.
	\begin{lemma}[$\boldsymbol{\mathcal{S}}$-procedure\cite{S-procedure}]\label{lemma-S-procedure}
		\emph{Let a quadratic function $f_j(\mathbf{x})$, $\mathbf{x}\in\mathbb{C}^{N\times 1}$, $\forall j\in\mathcal{J}=\{0,1,\cdots,J\}$, be defined as
			\begin{eqnarray}
				\label{C_Lemma1}
				f_j(\mathbf{x})=\mathbf{x}^{\mathrm H}\mathbf{A}_j\mathbf{x}+2{\rm Re}\{\mathbf{a}_j^{\mathrm H}\mathbf{x}\}+a_j, 
			\end{eqnarray}
			where $\mathbf{A}_j\in\mathbb{H}^N$ and $\mathbf{a}_j\in\mathbb{C}^{N\times 1}$.
			Then, the condition $\{f_j(\mathbf{x})\geq 0\}_{j=1}^J\Rightarrow f_0(\mathbf{x})\geq 0$ holds if and only if there exist $\upsilon_j\geq 0$, $\forall j$, such that}
		\begin{eqnarray}
			\Bigg[\begin{array}{cc}
				\mathbf{A}_0 & \mathbf{a}_0 \\
				\mathbf{a}_0^{\mathrm H} & a_0
			\end{array}\Bigg]-\sum\nolimits_{j=1}^{J}\upsilon_j\Bigg[\begin{array}{cc}
				\mathbf{A}_j & \mathbf{a}_j \\
				\mathbf{a}_j^{\mathrm H} & a_j
			\end{array}\Bigg] \succeq \mathbf{0}.
		\end{eqnarray}
	\end{lemma}
	
	In order to apply Lemma \ref{lemma-S-procedure} to constraint (\ref{robust_A_SCA}), we rewrite the channel uncertainties $\Lambda_{h,k}$ and $\Lambda_{G,k}$ as the following quadratic expressions:
	\begin{align}
		\mathbf{x}_k^{\rm H} \mathbf{C}_1   \mathbf{x}_k -\xi_{h,k}^2\leq 0,
		~~\mathbf{x}_k^{\rm H}  \mathbf{C}_2 \mathbf{x}_k -\xi_{G,k}^2\leq 0,~\forall k,
	\end{align}
	where 
	\begin{align}
			\mathbf{C}_1=\Bigg[\begin{array}{cc}
			\mathbf{I}_{N} & \mathbf{0} \\
			\mathbf{0} & \mathbf{0}
		\end{array}\Bigg], ~ \mathbf{C}_2=\Bigg[\begin{array}{cc}
			\mathbf{0} & \mathbf{0} \\
			\mathbf{0} & \mathbf{I}_{M N}
		\end{array}\Bigg].
	\end{align}

	Then, based on Lemma \ref{lemma-S-procedure}, constraint (\ref{robust_A_SCA}) holds if and only if there exist $\upsilon_{h,k},\upsilon_{G,k}\geq 0$, such that 
	\begin{align}
		\nonumber
		& \left[\begin{array}{cc} 
			\mathbf{A}_{k} \! + \!\upsilon_{h,k}\mathbf{C}_1 \! +\! \upsilon_{G,k} \mathbf{C}_2 & \mathbf{a}_{k} \\
			\mathbf{a}_{k}^{\mathrm{H}} &a_{k} \! -\! \mathcal{A}_k^{-1} \! - \! \upsilon_{h,k}\xi_{h,k}^2\! -\! \upsilon_{G,k}\xi_{G,k}^2 
		\end{array}\right] \\
	\label{robust_A_SCA_LMIs}
	 & \succeq \mathbf{0}, ~\forall k.
	\end{align}
	
	Similarly, using $\mathbf{H}\!=\!\widetilde{\mathbf{H}}\!+\! \triangle \mathbf{H}$, constraints (\ref{C_P_RF_robust}) and (\ref{C_P_O_robust}) are, respectively, recast as
	\setlength{\abovedisplayskip}{4pt}
	\setlength{\belowdisplayskip}{4pt}
	\begin{subequations}
		\label{C_P_ORF_robust}
		\begin{align}
			\label{C_P_O_robust_linear}
			&\!\!\!\!\!\!\!\!\mathbf{y}_{m}^{\rm H}\mathbf{B}\mathbf{y}_{m} \!+\!2{\rm Re}\{  \widetilde{\mathbf{y}}_{m}^{\rm H}\mathbf{B} \mathbf{y}_{m} \} \! + \! \widetilde{\mathbf{y}}_{m}^{\rm H} \mathbf{B} \widetilde{\mathbf{y}}_{m} \! +\! b_{m}\! \geq \!0, ~\Lambda_{H},\forall m, \!\!\!\!\!\!\!\!\\
			\label{C_P_RF_robust_linear}
			&\!\!\!\!\!\!\!\!\mathbf{y}^{\rm H}\mathbf{B}\mathbf{y}+2{\rm Re}\{  \widetilde{\mathbf{y}}^{\rm H}\mathbf{B} \mathbf{y} \}+ \widetilde{\mathbf{y}}^{\rm H} \mathbf{B} \widetilde{\mathbf{y}}+ b\leq 0, ~\Lambda_{H}, 
		\end{align}
	\end{subequations}
	where
	\setlength{\abovedisplayskip}{3pt}
	\setlength{\belowdisplayskip}{3pt}
	\begin{align}
		\nonumber
		&\mathbf{y}_{m}={\rm vec} (\mathbf{T}_m\triangle \mathbf{H}),
		~\widetilde{\mathbf{y}}_{m}={\rm vec} (\mathbf{T}_m\widetilde{\mathbf{H}}), \\
		\nonumber
		& \mathbf{B}=\mathbf{I}_M\otimes\big(\sum\nolimits_{k=1}^K\!\mathbf{f}_k\mathbf{f}_k^{\rm H}\big), ~b_{m}=(1-\alpha_{m})\sigma_1^2-\zeta_m,\\
		\nonumber
		&\mathbf{y}={\rm vec}(\boldsymbol{\Theta}\triangle  \mathbf{H}),
		~\widetilde{\mathbf{y}}={\rm vec}(\boldsymbol{\Theta}\widetilde{\mathbf{H}}), ~b=\sigma_1^2 {\rm  Tr}\big(\mathbf{\Theta}\mathbf{\Theta}^{\rm H}\big)  - \bar{\mathcal{W}}_c.
	\end{align}

   Based on $\lVert\triangle\mathbf{H}\lVert_F\leq\!\xi_{H}$, we obtain $\lVert {\rm vec} (\mathbf{T}_m\triangle \mathbf{H})\lVert\leq\! \frac{(1-\alpha_{m})\xi_{H}}{\sqrt{M}}$ and
	$ \lVert  {\rm vec} (\boldsymbol{\Theta} \triangle \mathbf{H})\lVert\leq \frac{\xi_{H}\Vert\boldsymbol{\Theta}\lVert_F }{\sqrt{M}}$.
	Therefore, we have
	\begin{align}
	\label{CSI_uncertainty_x}
	\mathbf{y}_{m}^{\rm H}\mathbf{y}_{m}-\frac{(1-\alpha_{m})^2\xi_{H}^2}{M}\leq 0,
	~~~\mathbf{y}^{\rm H}\mathbf{y}-\frac{\xi_{ H}^2\lVert\boldsymbol{\Theta}\lVert_F^2}{M}\leq 0.
	\end{align}
	According to (\ref{CSI_uncertainty_x}) and Lemma \ref{lemma-S-procedure}, with slack variables $\upsilon_{H,m}\geq 0$ and $\upsilon_{H}\geq 0$, constraints (\ref{C_P_O_robust_linear}) and (\ref{C_P_RF_robust_linear}) are transformed into the following linear matrix inequality (LMI) constraints:
	\begin{subequations}
		\begin{align}
		\nonumber
			&\left[\begin{array}{cc}
				\upsilon_{H,m}\mathbf{I}_{M N} & \mathbf{0} \\
				0 &b_m-\upsilon_{H,m} \frac{(1-\alpha_{m})^2\xi_{H}^2}{M}
			\end{array}\right] \\
			\label{C_P_O_robust_LMIs}
		  & +  \left[\begin{array}{c}
				\mathbf{I}_{M N} \\
				\widetilde{\mathbf{y}}_{m}^{\rm H}
			\end{array}\right]  \mathbf{B}  \left[\begin{array}{cc}
				\mathbf{I}_{M N} \ \ 
				\widetilde{\mathbf{y}}_{m}
			\end{array}\right] \succeq \mathbf{0}, ~\forall m, \\
			\nonumber
			&\left[\begin{array}{cc}
				\upsilon_{H}\mathbf{I}_{M N} & \mathbf{0} \\
				0 &- b-\upsilon_{H}\frac{\xi_{ H}^2\lVert\boldsymbol{\Theta}\lVert_F^2}{M} 
			\end{array}\right] \\
		 \label{C_P_RF_robust_LMIs}
		  &- \left[\begin{array}{c}
				\mathbf{I}_{M N}\\
				\widetilde{\mathbf{y}}^{\rm H}
			\end{array}\right]  \mathbf{B}  \left[\begin{array}{cc}
				\mathbf{I}_{M N} \  \ 
				\widetilde{\mathbf{y}}
			\end{array}\right] \succeq \mathbf{0}.
		\end{align}
	\end{subequations}

	Next, we consider the CSI uncertainties in $\Lambda_{h,k}$ and $\Lambda_{g,k}$ of constraint (\ref{C_B_robust}).
	By defining the matrix $\mathbf{F}_{-k}=\big[\mathbf{f}_1,\cdots,\mathbf{f}_{k-1},\mathbf{f}_{k+1},\cdots,\mathbf{f}_K\big]\in\mathbb{C}^{N\times (K-1)}$ and introducing a slack variable $\mathcal{D}_k$, constraint (\ref{C_B_robust}) is rewritten as
	\begin{subequations}
	\begin{align}
		\label{C_B_robust_recast_1}
		&\!\!\!\mathcal{B}_{k}\geq \lVert (\mathbf{h}_{k}^{\rm H}+\mathbf{v}^{\rm H}\mathbf{G}_{k})\mathbf{F}_{-k}\lVert^2 + \mathcal{D}_k +\sigma_0^2,~\Lambda_{h,k},\Lambda_{G}, \forall k, \!\!\!\\
		\label{C_B_robust_recast_2}
		&\!\!\!\mathcal{D}_k \geq \sigma_1^2\lVert\mathbf{g}_{k}^{\mathrm H}\boldsymbol{\Theta}\lVert^2, ~\Lambda_{g,k}, \forall k.\!\!\!
	\end{align}
   \end{subequations}
	Then, we adopt Schur's complement Lemma to equivalently recast constraints (\ref{C_B_robust_recast_1}) and (\ref{C_B_robust_recast_2}) as\cite{Schur's complement}
	\begin{subequations}
		\begin{align}
			\nonumber
			&\left[\begin{array}{cc}
				\mathcal{B}_{k}-\mathcal{D}_k-\sigma_0^2 &  (\mathbf{h}_{k}^{\rm H}+\mathbf{v}^{\mathrm{H}}\mathbf{G}_{k})\mathbf{F}_{-k}\\
				\mathbf{F}_{-k}^{\mathrm{H}}(\mathbf{h}_{k}+\mathbf{G}_{k}^{\mathrm{H}} \mathbf{v})   &\mathbf{I}_{K-1}
			\end{array}\right] \succeq \mathbf{0},  \\
		\label{C_B_robust_matrix-1-1}
		  &~\Lambda_{h,k}, \Lambda_{G}, \forall k,\\
			\label{C_B_robust_matrix-2-1}
			&\left[\begin{array}{cc}
				\mathcal{D}_k & \sigma_1\mathbf{g}_{k}^{\mathrm H}\boldsymbol{\Theta} \\
				\sigma_1(\mathbf{g}_{k}^{\mathrm H}\boldsymbol{\Theta})^{\rm H}  &\mathbf{I}_{M}
			\end{array}\right] \succeq \mathbf{0}, ~\Lambda_{g,k}, \forall k.
		\end{align}
	\end{subequations}
	We further insert $\mathbf{h}_k=\widetilde{\mathbf{h}}_k+ \triangle \mathbf{h}_k$, $\mathbf{G}_k=\widetilde{\mathbf{G}}_k+  \triangle \mathbf{G}_k$, and $\mathbf{g}_k=\widetilde{\mathbf{g}}_k+  \triangle \mathbf{g}_k$ into (\ref{C_B_robust_matrix-1-1}) and (\ref{C_B_robust_matrix-2-1}). 
	Then, constraints (\ref{C_B_robust_matrix-1-1}) and (\ref{C_B_robust_matrix-2-1}) are, respectively, reformulated as
		\begin{subequations}
		\begin{align}
			\label{C_B_robust_matrix-1-2}
			\nonumber
			&\left[\!\!\! \begin{array}{cc}
				\mathcal{B}_{k}\! -\! \mathcal{D}_k \! - \! \sigma_0^2 & (\widetilde{\mathbf{h}}_{k}^{\rm H} \! + \! \mathbf{v}^{\mathrm{H}}\widetilde{\mathbf{G}}_{k})\mathbf{F}_{-k}\\
				\mathbf{F}_{-k}^{\mathrm{H}}(\widetilde{\mathbf{h}}_{k} \! +\! \widetilde{\mathbf{G}}_{k}^{\mathrm{H}} \mathbf{v}) &\mathbf{I}_{K-1}
			\end{array}\!\!\! \right] \! +\!  \left[\!\!\! \begin{array}{c}
				\mathbf{0} \\
				\mathbf{F}_{-k}^{\mathrm{H}}
			\end{array}\!\!\! \right]\left[\!\!\begin{array}{ll}
				\triangle \mathbf{h}_{k} & \mathbf{0}
			\end{array}\!\!\right] \\
			\nonumber
		 & +\left[\begin{array}{c}
				\triangle \mathbf{h}_{k}^{\mathrm{H}} \\
				\mathbf{0}
			\end{array}\right]\left[\begin{array}{ll}
				\mathbf{0} & \mathbf{F}_{-k}
			\end{array}\right]	+\left[\begin{array}{c}
				\mathbf{0} \\
				\mathbf{F}_{-k}^{\mathrm{H}}
			\end{array}\right] \triangle \mathbf{G}_{k}^{\mathrm{H}}\left[\begin{array}{cc}
				\mathbf{v}& \mathbf{0}
			\end{array}\right] \\
			& +\left[\begin{array}{c}
				\mathbf{v}^{\mathrm{H}} \\
				\mathbf{0}
			\end{array}\right] \triangle \mathbf{G}_{k}\left[\begin{array}{cc}
				\mathbf{0} &\mathbf{F}_{-k}
			\end{array}\right]\succeq \mathbf{0} , ~\Lambda_{h,k}, \Lambda_{G},\forall k, \\
			\nonumber
			&\left[\!\!\begin{array}{cc}
				\mathcal{D}_k & \!\! \sigma_1\widetilde{\mathbf{g}}_{k}^{\mathrm H}\boldsymbol{\Theta} \\
				\sigma_1(\widetilde{\mathbf{g}}_{k}^{\mathrm H}\boldsymbol{\Theta})^{\rm H}  &\mathbf{I}_{M}
			\end{array}\!\! \right] + \left[\!\!\begin{array}{c}
				\mathbf{0} \\
				\sigma_1\boldsymbol{\Theta}^{\rm H}
			\end{array}\!\!\right] \left[\!\!\begin{array}{ll}
				\triangle \mathbf{g}_{k}  & \mathbf{0} \end{array}\!\!\right] \\
			\label{C_B_robust_matrix-2-2}
			& +  \left[\!\!\begin{array}{c}   	\mathbf{0}\\
				\triangle \mathbf{g}_{k}^{\mathrm{H}} 
				\end{array}\!\!\right]
				 \left[\!\!\begin{array}{ll}
				\mathbf{0} &\! \boldsymbol{\Theta}\sigma_1
			\end{array}\!\!\right]	\succeq \mathbf{0} , ~\Lambda_{g,k},\forall k.
		\end{align}
	\end{subequations}
  We observe that constraints (\ref{C_B_robust_matrix-1-2}) and (\ref{C_B_robust_matrix-2-2}) are still intractable due to the multiple complex valued uncertainties.
  Here, we transform them into a finite number of constraints by applying the following lemma.
	\begin{lemma}[General sign-definiteness\cite{General_sign_definiteness}]\label{lemma-sign}
		\emph{Given matrices $\mathbf{D}$ and $\{\mathbf{E}_j,\mathbf{F}_j\}_{j=1}^{J}$ with $\mathbf{D}=\mathbf{D}^{\rm H}$, the following semi-infinite LMI
		}
		\begin{eqnarray}
			\label{C_lemma2}
			\mathbf{D} \succeq \sum\nolimits_{j=1}^{J}\left(\mathbf{E}_j^{\mathrm{H}} \mathbf{G}_j \mathbf{F}_j+\mathbf{F}_j^{\mathrm{H}} \mathbf{G}_j^{\mathrm{H}} \mathbf{E}_j\right),  ~\left\|\mathbf{G}_j\right\|_{F} \leq \xi_j,~~\forall j,
		\end{eqnarray}
		\emph{
			holds if and only if there exist $\varpi_j\geq 0$, $\forall j$, such that
		}
		\begin{eqnarray}
			\left[\begin{array}{cccc}
				\mathbf{D}-\sum_{j=1}^{J} \varpi_{j} \mathbf{F}_{j}^{\mathrm{H}} \mathbf{F}_{j} & -\xi_{1} \mathbf{E}_{1}^{\mathrm{H}} & \cdots & -\xi_{J} \mathbf{E}_{J}^{\mathrm{H}} \\
				-\xi_{1} \mathbf{E}_{1} & \varpi_{1} \mathbf{I} & \cdots & \mathbf{0}\\
				\vdots & \vdots & \ddots & \vdots \\
				-\xi_{J} \mathbf{E}_{J} & \mathbf{0} & \cdots & \varpi_{J} \mathbf{I}
			\end{array}\right] \succeq \mathbf{0}.
		\end{eqnarray}
	\end{lemma}
	
	Let us take constraint (\ref{C_B_robust_matrix-1-2}) as an example.
	It is observed that constraint (\ref{C_B_robust_matrix-1-2}) can be recast by setting the parameters in Lemma \ref{lemma-sign} as follows:
	\begin{subequations}
		\label{Lemma2_coefficients}
		\begin{align}
			&J\!=\!2, ~\mathbf{D}\!=\!\left[\!\!\begin{array}{cc}
				\mathcal{B}_{k}\!-\!\mathcal{D}_k\!-\!\sigma_0^2 \!\!\!\!&(\widetilde{\mathbf{h}}_{k}^{\rm H}\!+\!\mathbf{v}^{\mathrm{H}}\widetilde{\mathbf{G}}_{k})\mathbf{F}_{-k}  \\
				\mathbf{F}_{-k}^{\mathrm{H}}(\widetilde{\mathbf{h}}_{k}\!+\!\widetilde{\mathbf{G}}_{k}^{\mathrm{H}} \mathbf{v})   \!\!\!\!&\mathbf{I}_{K-1}
			\end{array}\!\!\right], \\
			&\mathbf{G}_1\!=\!\left[\begin{array}{ll}
				\triangle \mathbf{h}_{k} &\mathbf{0}
			\end{array}\right],
			\mathbf{E}_1\!=\!-\left[\begin{array}{ll}
				\mathbf{0} & \mathbf{F}_{-k}
			\end{array}\right],
			~\mathbf{F}_1\!=\!\mathbf{I}_K, \\
			& \mathbf{G}_2\!=\!\triangle \mathbf{G}_{k}^{\mathrm{H}}, 
			~\mathbf{E}_2\!=\!-\left[\begin{array}{cc}
				\mathbf{0} & \mathbf{F}_{-k}
			\end{array}\right],
			~\mathbf{F}_2\!=\!\left[\begin{array}{cc}
				\mathbf{v} & \mathbf{0}
			\end{array}\right].
		\end{align}
	\end{subequations}
	Constraint (\ref{C_B_robust_matrix-1-2}) is then equivalently transformed into LMIs (\ref{C_B_robust_matrix-1_LMIs}) at the top of this page, where $\varpi_{h,k}\geq 0$ and $\varpi_{G,k}\geq 0$ are introduced slack variables. 
		\begin{figure*}[t]
		\begin{eqnarray}
			\label{C_B_robust_matrix-1_LMIs}
			\left[\begin{array}{cccc}
				\mathcal{B}_{k}-\mathcal{D}_k-\sigma_0^2-\varpi_{h,k}-\varpi_{ G,k}\sum_{m=1}^M{\alpha_{m}^2\beta_{m}} & (\widetilde{\mathbf{h}}_{k}^{\rm H}+\mathbf{v}^{\mathrm{H}}\widetilde{\mathbf{G}}_{k})\mathbf{F}_{-k}  &  \mathbf{0}&\mathbf{0}\\
				\mathbf{F}_{-k}^{\mathrm{H}}(\widetilde{\mathbf{h}}_{k}+\widetilde{\mathbf{G}}_{k}^{\mathrm{H}} \mathbf{v})  & (1-\varpi_{h,k} )\mathbf{I}_{K-1} & \xi_{h,k} \mathbf{F}_{-k}^{\mathrm{H}} &\xi_{G,k} \mathbf{F}_{-k}^{\mathrm{H}} \\
				\mathbf{0} & \xi_{h,k} \mathbf{F}_{-k} &\varpi_{h,k} \mathbf{I}_{N} &\mathbf{0}\\
				\mathbf{0}& \xi_{G,k} \mathbf{F}_{-k} & \mathbf{0} &\varpi_{G,k} \mathbf{I}_{N}
			\end{array}\right] \succeq \mathbf{0}, ~ \forall k.
		\end{eqnarray} 
		\hrulefill
		\vspace{-1mm}
	\end{figure*}
	Similarly, given the introduced slack variable $\varpi_{g,k}\geq 0$, the equivalent LMIs of constraint (\ref{C_B_robust_matrix-2-2}) are obtained as 
	\begin{eqnarray}
		\label{C_B_robust_matrix-2_LMIs}
		\left[\begin{array}{ccc}
			\mathcal{D}_k-\varpi_{g,k}& \sigma_1\widetilde{\mathbf{g}}_{k}^{\mathrm H}\boldsymbol{\Theta}
			& \mathbf{0} \\
			\sigma_1(\widetilde{\mathbf{g}}_{k}^{\mathrm H}\boldsymbol{\Theta})^{\rm H}   &  (1-\varpi_{g,k} )\mathbf{I}_{M} & \xi_{g,k} \sigma_1 \boldsymbol{\Theta}^{\rm H} \\
			\mathbf{0} & \xi_{g,k}\sigma_1 \boldsymbol{\Theta}&\varpi_{g,k} \mathbf{I}_{M}
		\end{array}\right] \succeq \mathbf{0},  ~\forall k.
	\end{eqnarray}

	As a result, by replacing the original constraints (\ref{C_A_robust})-(\ref{C_P_O_robust}) with the LMI constraints (\ref{robust_A_SCA_LMIs}), (\ref{C_B_robust_matrix-1_LMIs}), (\ref{C_B_robust_matrix-2_LMIs}), (\ref{C_P_O_robust_LMIs}), and (\ref{C_P_RF_robust_LMIs}), respectively, problem (\ref{P0-robust-1}) is reformulated as
	\begin{subequations}
		\label{P0-robust-2}
		\begin{eqnarray}
			&\underset{\mathbf{f}_k,\boldsymbol{\Theta},\Delta,\Delta_0}{\max}  &\sum\nolimits_{k=1}^KQ_{k}\\
			\label{C-LMI-coefficients-1}
			&\operatorname{s.t.} &\upsilon_{h,k},\upsilon_{G,k},\upsilon_{H,m},\upsilon_{H}, \geq 0 , ~\forall k,\forall m, \\
			\label{C-LMI-coefficients-2}
			&& \varpi_{h,k},\varpi_{G,k},\varpi_{g,k}\geq 0, ~\forall k,\\
			&& {\rm (\ref{C_transmit beamforming}), (\ref{C-MF-RIS}), (\ref{C_energy-3}), (\ref{C_Q_lb}), (\ref{robust_A_SCA_LMIs})}, \\
			&&{\rm (\ref{C_P_O_robust_LMIs}), (\ref{C_P_RF_robust_LMIs}), (\ref{C_B_robust_matrix-1_LMIs}), (\ref{C_B_robust_matrix-2_LMIs}) 
			},
		\end{eqnarray}
	\end{subequations}
	where $\Delta_0=\{\upsilon_{h,k},\upsilon_{G,k},\upsilon_{H,m},\upsilon_{H},\varpi_{h,k},\varpi_{G,k},\varpi_{g,k},\mathcal{D}_k|\forall k, $ $\forall m\}$ represents the slack variable set.
	The resulting multi-variate optimization problem (\ref{P0-robust-2}) can be solved using the typical AO method. The details for updating each variable are given in the next subsection.
	
	\subsection{Joint Design of Transmit Beamforming and MF-RIS Coefficients}
	\subsubsection{Optimizing $\mathbf{f}_k$ with given $\boldsymbol{\Theta}$}
	With fixed MF-RIS coefficient $\boldsymbol{\Theta}$, the transmit beamforming optimization problem under imperfect CSI is written as 
	\begin{subequations}
		\label{P-f-robust-1}
		\begin{eqnarray}
			&\underset{\mathbf{f}_k,\Delta,\Delta_0}{\max}  &\sum\nolimits_{k=1}^KQ_{k}\\
			&\operatorname{s.t.} & {\rm (\ref{C_transmit beamforming}), (\ref{C_energy-3}), (\ref{C_Q_lb}), (\ref{robust_A_SCA_LMIs}), (\ref{C_P_O_robust_LMIs})  }, \\
			&&{\rm (\ref{C_P_RF_robust_LMIs}), (\ref{C_B_robust_matrix-1_LMIs}), (\ref{C_B_robust_matrix-2_LMIs}), (\ref{C-LMI-coefficients-1}), (\ref{C-LMI-coefficients-2})}.
		\end{eqnarray}
	\end{subequations}
    Problem (\ref{P-f-robust-1}) is a convex SDP, and thus can be solved efficiently via CVX\cite{CVX}.
    
	\subsubsection{Optimizing $\boldsymbol{\Theta}$ with given $\mathbf{f}_k$}
	Given the transmit beamforming vector $\mathbf{f}_k$, the MF-RIS coefficient optimization problem is formulated as
	\begin{subequations}
		\label{P-passive-robust-1}
		\begin{eqnarray}
			&\!\!\!\!\!\!\underset{\mathbf{v},\Delta,\Delta_0}{\max}  &\sum\nolimits_{k=1}^KQ_{k}\\
			\label{C-passive-robust-v}
			&\!\!\!\!\!\!\operatorname{s.t.} &\left[\mathbf{v}\right]_m=\alpha_{m}\sqrt{\beta_m} e^{j\theta_m}, ~\theta_m\in[0,2\pi), ~\forall m,\\
			\label{C-passive-robust-alpha}
			&& \alpha_m\in\{0,1\}, ~\beta_{m}\in[0,\beta_{\rm max}], ~\forall m,\\
			\label{C-passive-robust-beta-other}
			&&{\rm (\ref{C_energy-3}), (\ref{C_Q_lb}), (\ref{robust_A_SCA_LMIs}),(\ref{C_P_O_robust_LMIs}), (\ref{C_P_RF_robust_LMIs}),} \\
			&&{\rm (\ref{C_B_robust_matrix-1_LMIs}), (\ref{C_B_robust_matrix-2_LMIs}), (\ref{C-LMI-coefficients-1}), (\ref{C-LMI-coefficients-2}).
			}
		\end{eqnarray}
	\end{subequations}

	The difficulties of solving (\ref{P-passive-robust-1}) lie in the non-convex LMIs (\ref{C_P_O_robust_LMIs}) and (\ref{C_B_robust_matrix-1_LMIs}), the highly-coupled unit-modulus constraint (\ref{C-passive-robust-v}), and the binary constraint in (\ref{C-passive-robust-alpha}).
   By replacing the non-convex terms $\upsilon_{H,m}(1-\alpha_{m})^2$ in (\ref{C_P_O_robust_LMIs}) and ${\alpha_{m}^2\beta_{m}}$ in (\ref{C_B_robust_matrix-1_LMIs}) with their FTSs $(1\!-\!\alpha_{m}^{(\ell)})(\upsilon_{H,m}\!-\!\upsilon_{H,m}\alpha_{m}^{(\ell)}\!-\!2 \upsilon_{H,m}^{(\ell)}\alpha_{m}\!+\!2 \upsilon_{H,m}^{(\ell)}\alpha_{m}^{(\ell)})$ and $ 2(\alpha_{m}\!-\!\alpha_{m}^{(\ell)})\alpha_{m}^{(\ell)}\beta_{m}^{(\ell)}+(\alpha_{m}^{(\ell)})^2\beta_{m}$, respectively, LMIs (\ref{C_P_O_robust_LMIs}) and (\ref{C_B_robust_matrix-1_LMIs}) are recast as their convex approximations
    (\ref{C_P_O_robust_LMIs})$'$ and (\ref{C_B_robust_matrix-1_LMIs})$'$, where $\{\upsilon_{H,m}^{(\ell)}, \alpha_{m}^{(\ell)},\beta_{m}^{(\ell)}\}$ is the feasible point in the $\ell$-th iteration.
   The expressions for (\ref{C_P_O_robust_LMIs})$'$ and (\ref{C_B_robust_matrix-1_LMIs})$'$ are omitted here for brevity.
   
	Similar to the transformation of constraint (\ref{C_passive_rank_2-relax}), we here adopt the penalty function-based method to address constraint (\ref{C-passive-robust-v}).
	By introducing an auxiliary variable $\eta_m=\alpha_{m}^2\beta_{m}$, we obtain the equivalent form of (\ref{C-passive-robust-v}) as
	\begin{align}
		\label{C-passive-robust-v-relax}
		\left[\mathbf{v}\right]_{m}=\sqrt{\eta_m}e^{j\theta_m}, 
		~\eta_m\leq \alpha_{m}^2\beta_{m},
		~\alpha_{m}^2\beta_{m} \leq \eta_m, ~\forall m.
	\end{align}
	With the aid of an auxiliary variable set $\mathbf{e}=\{e_m|\forall m\}$, satisfying $e_m=\left[\mathbf{v}\right]_{m}^{\ast}\left[\mathbf{v}\right]_{m}$, the unit-modulus constraint $\left[\mathbf{v}\right]_{m}=\sqrt{\eta_m}e^{j\theta_m}$ is linearized as $e_m\leq \left[\mathbf{v}\right]_{m}^{\ast}\left[\mathbf{v}\right]_{m}\leq e_m$.
	Following the FTS, we further approximate the non-convex part $e_m\leq \left[\mathbf{v}\right]_{m}^{\ast}\left[\mathbf{v}\right]_{m}$ by $e_m\leq 2 {\rm Re} \left\{ \left[\mathbf{v}\right]_{m}^{\ast} \left[\mathbf{v}^{(\ell)}\right]_{m}\right\}-\left[\mathbf{v}^{(\ell)}\right]_m^{\ast}\left[\mathbf{v}^{(\ell)}\right]_m$.
    In Section \ref{MF-RIS Coefficient Design}, we showed how to deal with the non-convex constraints $\eta_m\leq \alpha_{m}^2\beta_{m}$ and $\alpha_{m}^2\beta_{m} \leq \eta_m$, and the binary constraint in (\ref{C-passive-robust-alpha}).
    Therefore, by introducing a slack variable set $\mathbf{q}=\{q_m,\bar{q}_m|\forall  m\}$, problem (\ref{P-passive-robust-1}) is transformed into
	\begin{subequations}
		\label{P-passive-robust-2}
		\begin{eqnarray}
			&\!\!\!\!\!\!\!\!\!\!\!\!\!\!\!\!\underset{\mathbf{v},\Delta,\Delta_0, \boldsymbol{\eta}, \mathbf{d},\mathbf{e},\mathbf{q}}{\max}  &\!\!\!\!\! \sum\nolimits_{k=1}^K \! Q_{k} \!\! - \!\! \rho \sum\nolimits_{m=1}^{M}(d_m \! + \!\bar{d}_m \! + \! q_m \! + \! \bar{q}_m)\\
			&\!\!\!\!\!\!\!\!\!\!\!\!\!\operatorname{s.t.} &  \!\!\!  \left[\mathbf{v}\right]_{m}^{\ast}\left[\mathbf{v}\right]_{m}\leq e_m+q_m,~\forall m,\\
			\nonumber
			&&\!\!\! 2 {\rm Re} \left\{ \left[\mathbf{v}\right]_{m}^{\ast} \left[\mathbf{v}^{(\ell)}\right]_{m}\right\}-\left[\mathbf{v}^{(\ell)}\right]_m^{\ast}\left[\mathbf{v}^{(\ell)}\right]_m \\
			&& \!\!\! \geq e_m-\bar{q}_m,~\forall m,\\
			&& \!\!\! e_m,\beta_m \in[0,\beta_{\max}], ~\forall m, \\
			&&\!\!\! {\rm (\ref{C_energy-3}), (\ref{C_Q_lb}),  (\ref{C-passive-alpha}){\text -}(\ref{C-passive-alpha-beta}), (\ref{robust_A_SCA_LMIs})},\\
			&&\!\!\! {\rm (\ref{C_P_RF_robust_LMIs}), (\ref{C_B_robust_matrix-2_LMIs}), (\ref{C-LMI-coefficients-1}), (\ref{C-LMI-coefficients-2}), (\ref{C_P_O_robust_LMIs})^{'}, (\ref{C_B_robust_matrix-1_LMIs})^{'}.
			}
		\end{eqnarray}
	\end{subequations}
	Problem (\ref{P-passive-robust-2}) is a convex SDP, which can be solved efficiently via CVX\cite{CVX}.
	The algorithm for solving problem (\ref{P-passive-robust-1}) is similar to Algorithm \ref{Algorithm_passive}, and thus is omitted for simplicity.
	
	Next, we analyze the computational complexity of our robust scheme.
	It is observed that both resulting problems (\ref{P-f-robust-1}) and (\ref{P-passive-robust-1}) involve LMI, second-order cone, and linear constraints, and thus can be solved efficiently via the interior point method\cite{Schur's complement}. 
     According to the general complexity expression given in \cite{Robust-TSP},
	  the complexity of solving problem (\ref{P-f-robust-1}) and problem (\ref{P-passive-robust-1}) is given by
	 $\mathcal{O}_{\mathbf{f}}=	\mathcal{O}\Big(\sqrt{(K(s_1+s_2+s_3)+(M+1)s_4)} n_1\big(n_1^2+n_1(K(s_1^2+s_2^2+s_3^2)+(M+1)s_4^2    )+K(s_1^3+s_2^3+s_3^3)+(M+1)s_4^3\big)\Big)$ and  $\mathcal{O}_{\boldsymbol{\Theta}}=\mathcal{O}\Big(\sqrt{(K(s_1+s_2+s_3)+(M+1)s_4+4M)}$ $n_2\big(n_2^2+n_2(K(s_1^2+s_2^2+s_3^2)+(M+1)s_4^2    )+K(s_1^3+s_2^3+s_3^3)+(M+1)s_4^3+2n_2M\big)\Big)$, respectively, where $n_1=NK$, $n_2=2M$, $s_1=N(M+1)+1$, $s_2=2N+K$, $s_3=M+2$, and $s_4=MN+1$.
    Similar to the perfect CSI case in Section \ref{Perfect CSI case}, the convergence of the robust beamforming scheme can be proved and thus is omitted here for simplicity.

	\section{Numerical Results} \label{Numerical Results}	
\begin{figure}[t]
  	\centering
  		\centering
  		  \includegraphics[width=3.05 in]{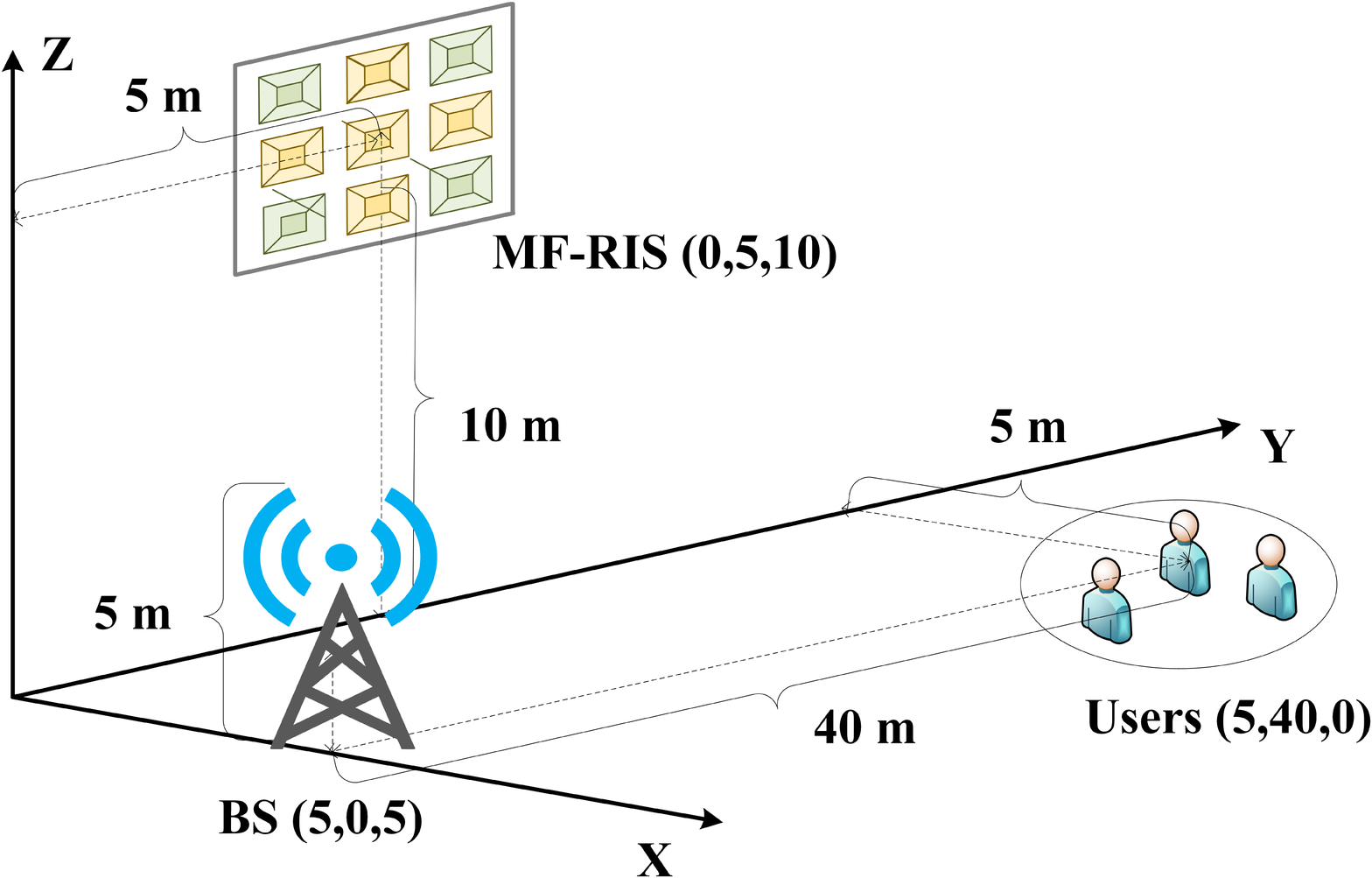}
  		\vspace{-2mm}
  		\caption{Simulation setup of the MF-RIS-aided communication network.}
  		\label{simulation_model}
  		\vspace{-1mm}
  	\end{figure}

	In this section, numerical results are provided to evaluate the performance of the considered MF-RIS-aided wireless network.
	As shown in Fig. \ref{simulation_model}, the BS and the MF-RIS are located at $(5,0,5)$ and $(0,5,10)$ m, respectively.
	We assume that $K=3$ users are randomly distributed in a circle centered at $(5,40,0)$ m with the radius of $4$ m.
	All channels are modeled by Rician fading. 
	We define the maximum normalized estimation error as $\kappa_{h,k}=\frac{\xi_{h,k}}{\lVert\widetilde{\mathbf{h}}_k\lVert}$, $\kappa_{g,k}=\frac{\xi_{g,k}}{\lVert\widetilde{\mathbf{g}}_k\lVert}$, and
	$\kappa_{H}=\frac{\xi_{H}}{\lVert\widetilde{\mathbf{H}}\lVert_F}$\cite{Secure_Wang_JSTSP}.
	Then, following the Cauchy-Schwarz inequality and $\mathbf{G}_k={\rm diag}(\mathbf{g}_k^{\rm H})\mathbf{H}$, we obtain $\xi_{G,k}=\xi_{H}\lVert {\rm diag}(\widetilde{\mathbf{g}}_k^{\rm H})\lVert_F + \xi_{g,k}\lVert \widetilde{\mathbf{H}}\lVert_F+\xi_{g,k}\xi_{H}= \frac{(\xi_{g,k}\xi_{H})(\kappa_{H}+\kappa_{g,k}+\kappa_{H}\kappa_{g,k})}{\kappa_{H}\kappa_{g,k}}$.
   Unless otherwise specified, we set $\kappa_{h,k}^2\!=\!\kappa_{g,k}^2\!=\!\kappa_{H}^2\!=\!0.1$.
   More simulation settings are listed in Table \ref{Simulation parameters}.
   For comparison, we consider the self-sustainable RIS\cite{sustainable-RIS-TCOM} and reflecting-only RIS\cite{Throughput_NOMA_TWC} as benchmarks.
    
    \begin{table*}[t]\small
 	\centering
 	\renewcommand{\arraystretch}{1.2}
 	\caption{Simulation parameters}
 	\vspace{-2mm}
 	\label{Simulation parameters}
 	\scalebox{0.93}{
 		\begin{tabular}{|c|c|}
 			\hline  
 			\bfseries Parameter & \bfseries Value\\
 			\hline
 			Path loss at the reference distance of $1$ m & $-20$ dB\cite{sustainable-RIS-TC-WPT}\\
 			\hline 
 			Path loss exponents of BS-RIS, BS-user, and RIS-user links & $2.2$, $2.8$, $2.6$\\
 			\hline
 			Rician factors of BS-RIS, BS-user, and RIS-user links & $3$ dB\\
 			\hline
 			Noise power at users and the RIS & $\sigma_0^2=\sigma_1^2=-70$ dBm\\
 			\hline 
 			Energy harvesting and power consumption parameters & \tabincell{c}{$\xi=1.1$, $P_b=1.5$ mW, $P_{\rm DC}=0.3$ mW, $P_{\rm C}=2.1$ $\mu$W\cite{sustainable-RIS-TCOM},\\ $Z=24$ mW,  $a=150$, $q=0.014$\cite{WCL-Nonlinear-cost reduction}} \\
 			\hline
 			Other parameters & $N=4$, $\rho^{(0)}=10^{-3}$, $\varepsilon=10$\cite{Secure_Wang_JSTSP} \\
 			\hline
 	\end{tabular}}
 \end{table*}       
    
       \begin{figure}[t]
    	\centering
          \includegraphics[width=3.05in]{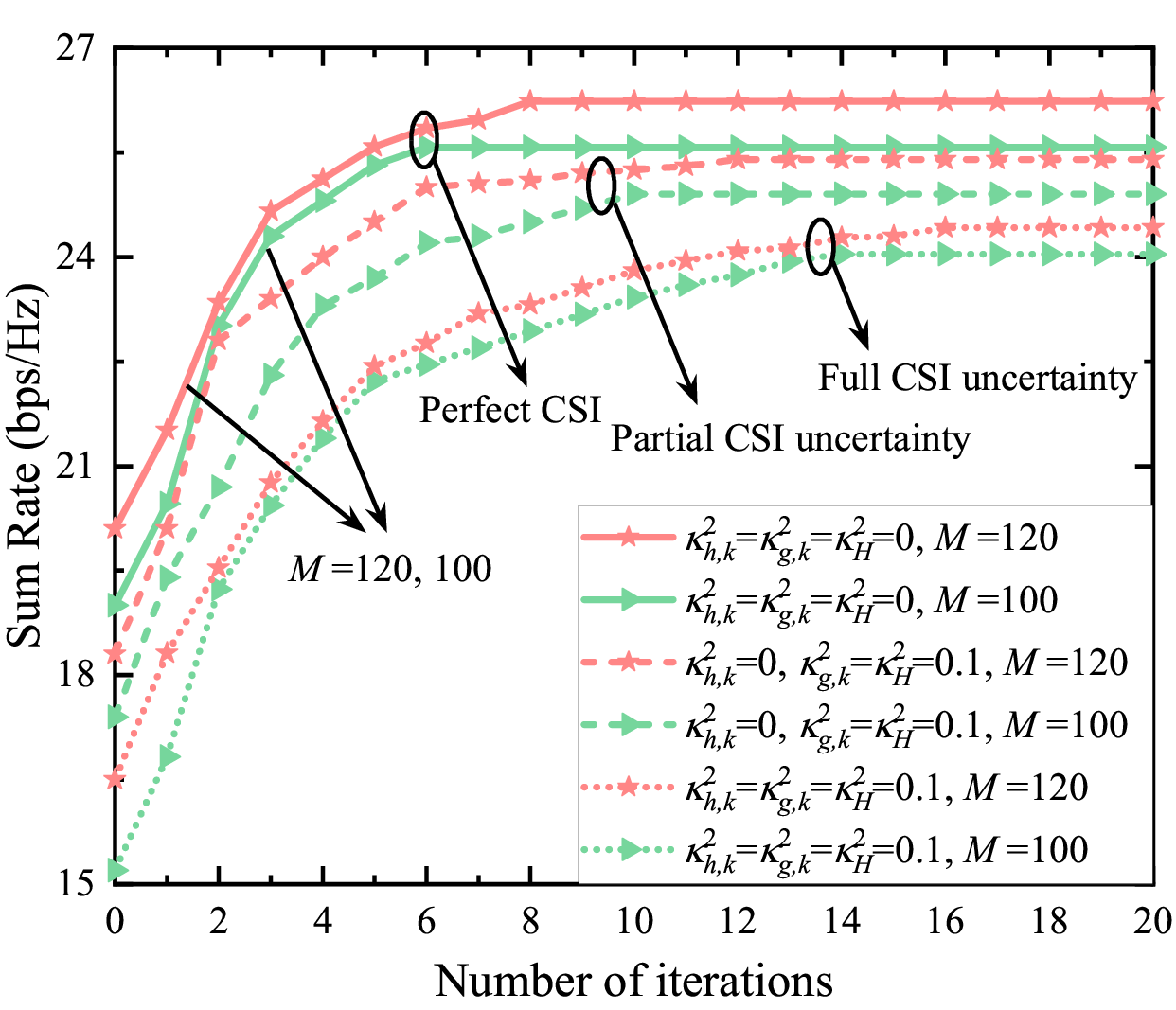}
           \vspace{-2mm}
          \caption{Convergence behavior of the proposed algorithm under different numbers of elements and different CSI setups, where $P_{\rm BS}^{\max}=36$ dBm and $\beta_{\max}=16$ dB.}
         \label{Convergence}
         \vspace{-1mm}
    \end{figure}

Fig. \ref{Convergence} illustrates the convergence behavior of the proposed algorithm with different CSI setups and different numbers of RIS elements.
It is observed that the proposed algorithm converges rapidly, e.g., $18$ iterations are sufficient for it to converge.
We notice that the convergence speed of the proposed algorithm with more elements is slightly slower than that with fewer elements.
This is because both the dimension of the solution space and the number of constraints increase with $M$, and thus increase the complexity of solving problems (\ref{P0}) and  (\ref{P0-robust-1}).
Meanwhile, Fig. \ref{Convergence} shows that the robust algorithm under full CSI uncertainty requires more iterations than that under partial CSI uncertainty, while the algorithm under perfect CSI converges fastest.
The reason is that the CSI uncertainty error increases the dimension of LMIs, which in turn increases the computational complexity of the proposed algorithm.

    \begin{figure}[t]
    	\centering
    		\includegraphics[width=3.05in]{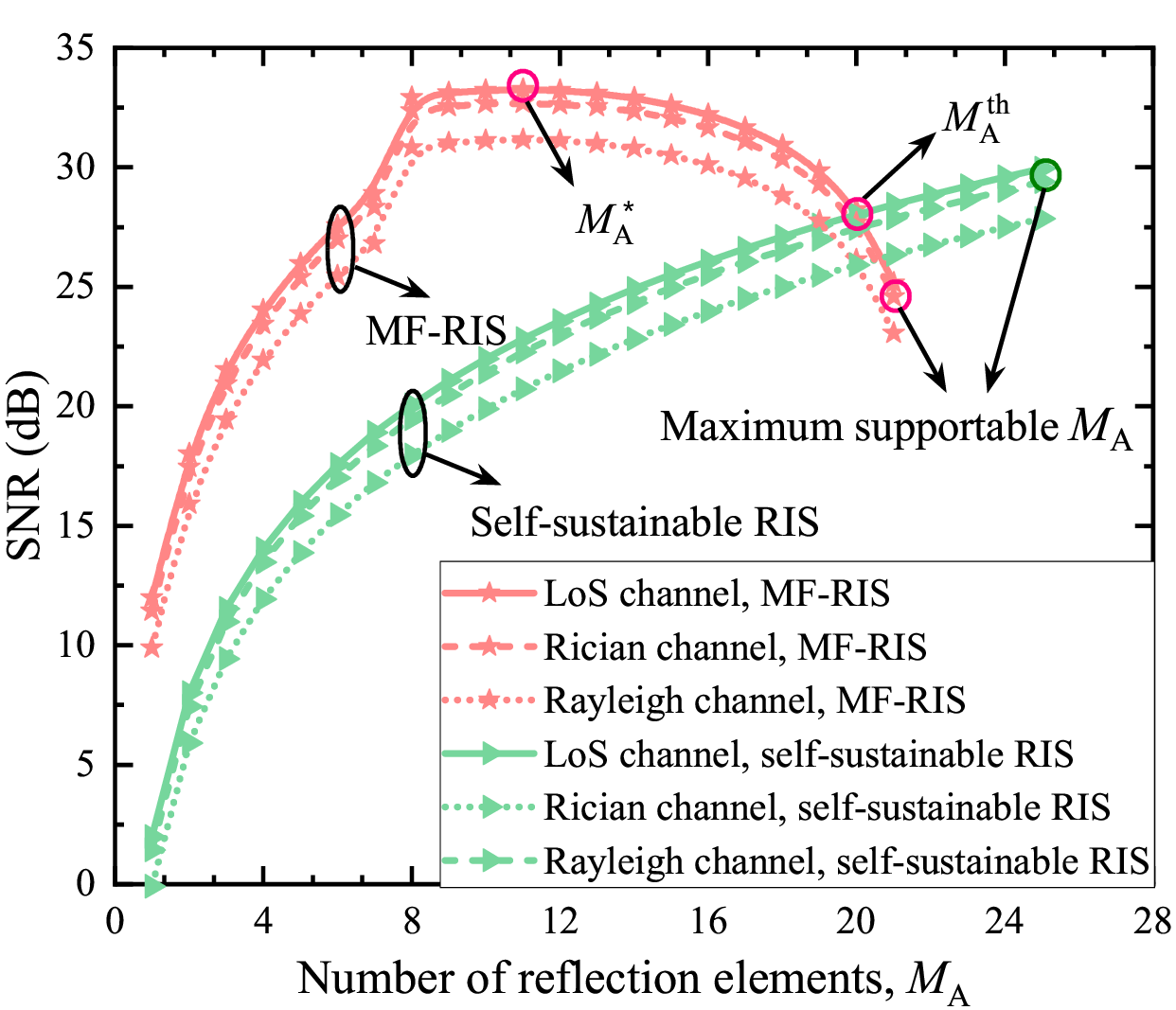}
    		\vspace{-2mm}
    		\caption{SNR versus $M_{\rm A}$ under different schemes and different channels, where the system model and parameter settings are the same as Section \ref{Analysis results}, and the Rician factor is $3$ dB.}
    		\label{M_H}
    		\vspace{-1mm}
    	 \end{figure}
        \begin{figure}[t]
    	\centering
    	\includegraphics[width=3.05in]{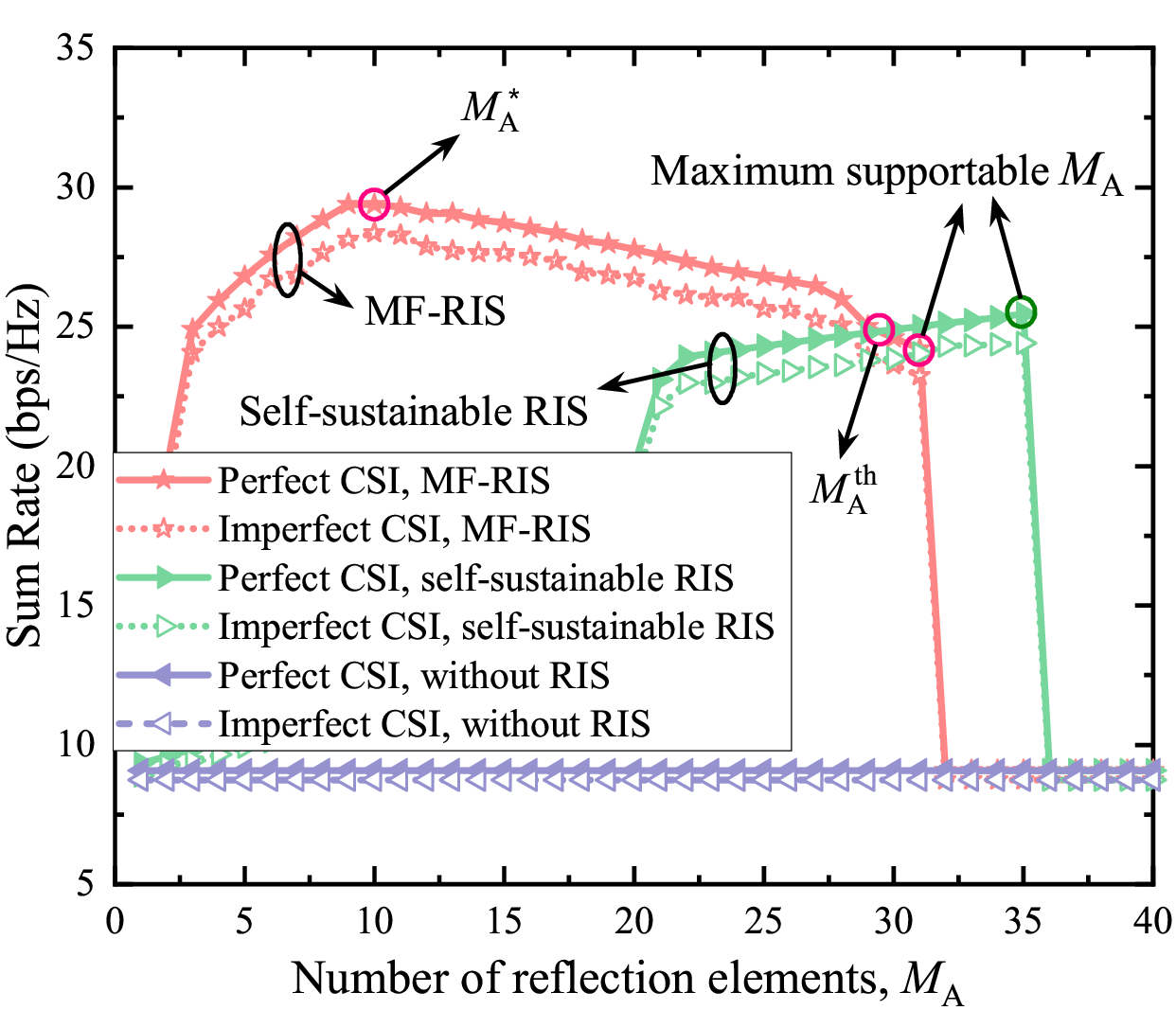}
    	\vspace{-2mm}
    	\caption{SR versus $M_{\rm A}$ under different schemes and different CSI setups, where $P_{\rm BS}^{\max}=40$ dBm and $M=130$. }
    	\label{M_H_MISO}
    	\vspace{-1mm}
    \end{figure}
   
   To verify the theoretical results in Section \ref{Analysis results}, we depict the achievable SNR and the achievable SR versus $M_{\rm A}$ in Figs. \ref{M_H} and \ref{M_H_MISO}, respectively.
   Specifically, Fig. \ref{M_H} is based on the single-user SISO system considered in Section \ref{Analysis results}, where the SNR values of the MF-RIS and self-sustainable RIS schemes under LoS channels are calculated using (\ref{MF-RIS-SNR-2}) and (\ref{SNR-SE-RIS}), respectively, while the numerical results of Rician and Rayleigh channels are obtained by averaging over $2000$ channel realizations.
   In contrast, Fig. \ref{M_H_MISO} is based on the simulation settings in Fig. \ref{simulation_model} and Table \ref{Simulation parameters}, where RIS-aided multi-user MISO systems with Rician channels are considered.
   It is observed that the curves under different channels (i.e., LoS, Rician, and Rayleigh channels), different CSI setups (i.e., perfect and imperfect CSI), and different numbers of transmit antennas and users (i.e., single-user SISO and multi-user MISO systems) exhibit similar trends, validating that our theoretical results can be used to guide the system design for the more general cases.
   With the increase of $M_{\rm A}$, the SNR and the SR for the MF-RIS first increase and then decrease after reaching $M_{\rm A}^{\star}$, which agrees well with our analysis in Section \ref{Analysis results}.
   This result characterizes the trade-off between $M_{\rm A}$ and $M_{\rm H}$ due to the fixed $M$, and the trade-off between $M_{\rm  A}$ and the amplification power due to the limited available power at the MF-RIS.
   Specifically, when $M_{\rm A}\leq M_{\rm A}^{\star}$, the available power at the MF-RIS with a large $M_{\rm H}$ is adequate, and thus the MF-RIS can benefit more from the increasing amplification gain brought by increasing $M_{\rm A}$.
   However, when $M_{\rm A}\geq M_{\rm A}^{\star}$, the available power at the MF-RIS is limited by the reduced $M_{\rm H}$.
   Meanwhile, a larger $M_{\rm A}$ introduces greater amplifier, phase shifter, and thermal noise power consumption, reducing the available amplification power, which makes the MF-RIS suffer more from the increased $M_{\rm A}$.
   
   Figs. \ref{M_H} and \ref{M_H_MISO} show that the achievable SNR and the achievable SR of the self-sustainable RIS increase with increasing $M_{\rm A}$.
   This is because when increasing $M_{\rm A}$, unlike the MF-RIS power constraint (\ref{analysis-problem-1-C2}) which reduces the available amplification power, the self-sustainable RIS power constraint (\ref{power-DF-RIS}) only limits its maximum supportable $M_{\rm A}$.
   Besides, as revealed in our theoretical results, the MF-RIS performs better than the self-sustainable RIS when $M_{\rm  A} \leq M_{\rm A}^{\rm th}$, but worse when $M_{\rm  A} > M_{\rm A}^{\rm th}$.
    This is because in this power limited case, the small amplification gain provided by the MF-RIS is outweighed by the adverse effect of its amplifier and thermal-noise power consumption on overall SNR performance.
   In addition, we notice that when $M_{\rm A}$ exceeds the maximum supportable value, the MF-RIS and the self-sustainable RIS may even fail to sustain themselves due to insufficient harvested energy.
   These results indicate that element allocation is crucial to improve the achievable performance of the proposed MF-RIS.
   Therefore, in Sections \ref{Perfect CSI case} and \ref{Imperfect CSI case}, a flexible element allocation model (by optimizing the mode indicator $\alpha_{m}$) is adopted to provide additional degrees of freedom for throughput improvement. 

	\begin{figure}[t]
		\centering
       \includegraphics[width=3.05in]{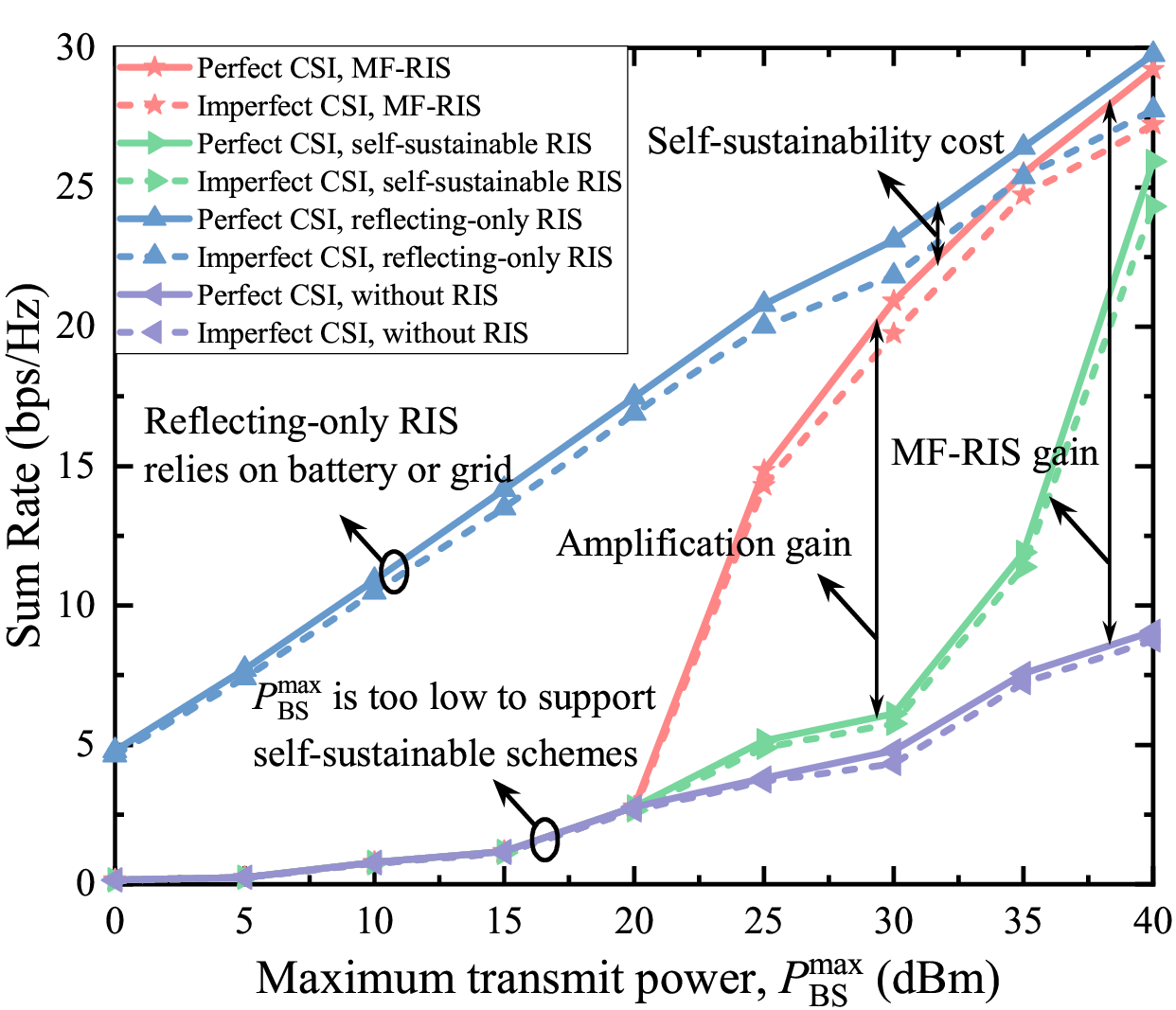}
         \vspace{-2mm}
         \caption{SR versus $P_{\rm BS}^{\rm max}$ under different schemes and different CSI setups, where $M=120$ and $\beta_{\max}=16$ dB.}
          \label{Pmax}
          	\vspace{-1mm}
    \end{figure}
    
    Fig. \ref{Pmax} shows the achievable SR versus $P_{\rm BS}^{\max}$ under different schemes and various CSI setups.
    We observe that when $P_{\rm BS}^{\max}$ is small, the battery- or grid-powered reflecting-only RIS can achieve satisfactory SR gain compared to the without RIS scheme, while the SR gains achieved by self-sustainable schemes are almost negligible.
    This is because at low power, only very limited power can be harvested by self-sustainable RIS schemes, which cannot even support their normal operation.
    However, the achievable SR values of the proposed MF-RIS scheme greatly exceed those of self-sustainable RIS and without RIS when $P_{\rm BS}^{\max}$ is moderate.
    Specifically, for the perfect CSI case with $P_{\rm BS}^{\max}=35$ dBm, the MF-RIS scheme is able to provide up to 114\% and 237\% higher SR than the self-sustainable RIS and without RIS counterparts, respectively.
    This reveals that the introduction of signal amplification can effectively alleviate the double-fading effect, thereby significantly improving the SR of all users.
    Additionally, we observe that the achievable SR of MF-RIS and self-sustainable RIS is inferior to that of reflecting-only RIS.
    This is because both the MF-RIS and self-sustainable RIS need to sacrifice part of their elements for energy harvesting to maintain self-sustainability. 
    In contrast, the reflecting-only RIS allows all elements to reflect the incident signal and enhance the desired reception.
    However, the self-sustainability cost of the self-sustainable RIS and MF-RIS decreases significantly with an increased $P_{\rm BS}^{\max}$, due to the fact that the elements operating in H mode can harvest more energy when the RF power is high.
    Especially for the MF-RIS, the performance gap between it and the reflecting-only RIS is negligible, which confirms the effectiveness of the proposed MF-RIS architecture.
    
\begin{figure}[t]
    	\centering
    	\includegraphics[width=3.05in]{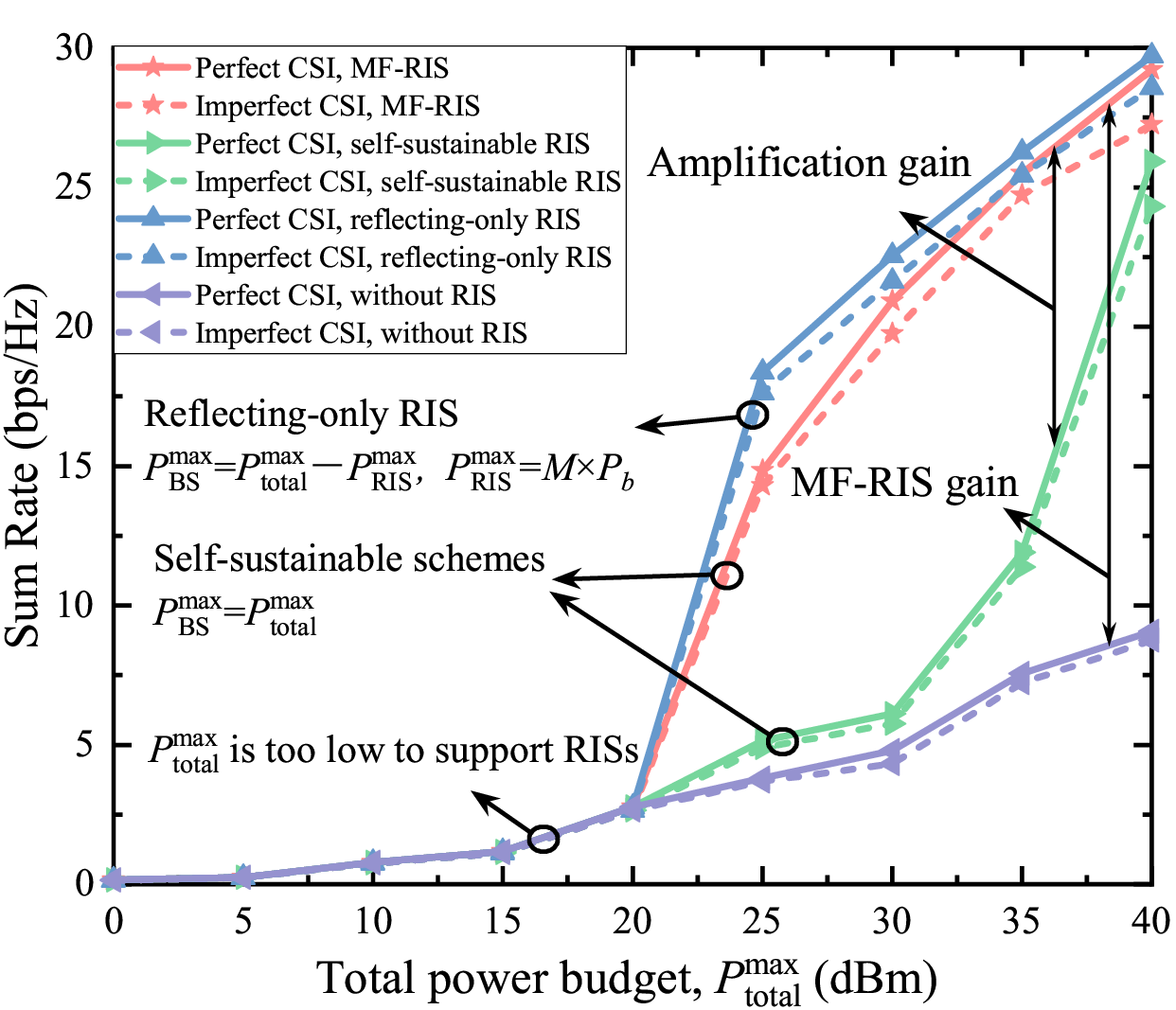}
    	\vspace{-2mm}
    	\caption{SR versus $P_{\rm total}^{\rm max}$ under different schemes and different CSI setups, where $M=120$ and $\beta_{\max}=16$ dB.}
    	\label{Ptotal}
    	\vspace{-1mm}
    \end{figure}

    Considering that the power consumption of the reflecting-only RIS is ignored in Fig. \ref{Pmax}, for fair comparison, we characterize the achievable SR versus the total power consumption $P_{\rm total}^{\max}$ under different schemes in Fig. \ref{Ptotal}.
    Specifically, $P_{\rm total}^{\max}=P_{\rm BS}^{\max}$ holds for MF-RIS, self-sustainable RIS, and without RIS schemes, while $P_{\rm total}^{\max}=P_{\rm BS}^{\max}+MP_b$ holds for the reflecting-only RIS scheme.
    Since its own power consumption is taken into account, the utility of the reflecting-only RIS at low power is as limited as that of self-sustainable schemes. 
    Nevertheless, when further increasing $P_{\rm total}^{\max}$, both the MF-RIS and reflecting-only RIS can provide considerable signal enhancement.
    It is seen that the proposed MF-RIS achieves slightly lower SR than the reflecting-only RIS at high $P_{\rm total}^{\max}$.
    This is because the inevitable power loss during energy harvesting and signal amplification is taken into account by the MF-RIS, while the reflecting-only RIS assumes an ideal lossless signal reflection and power supply process.
    However, the acceptable performance gap between the two shows that the proposed MF-RIS is a promising self-sustainable RIS architecture, especially for remote areas where it is difficult to lay power grids and manually replace batteries.
    
      \begin{figure}[t]
    		\centering
    		\vspace{2mm}
    		\includegraphics[width=3.05in]{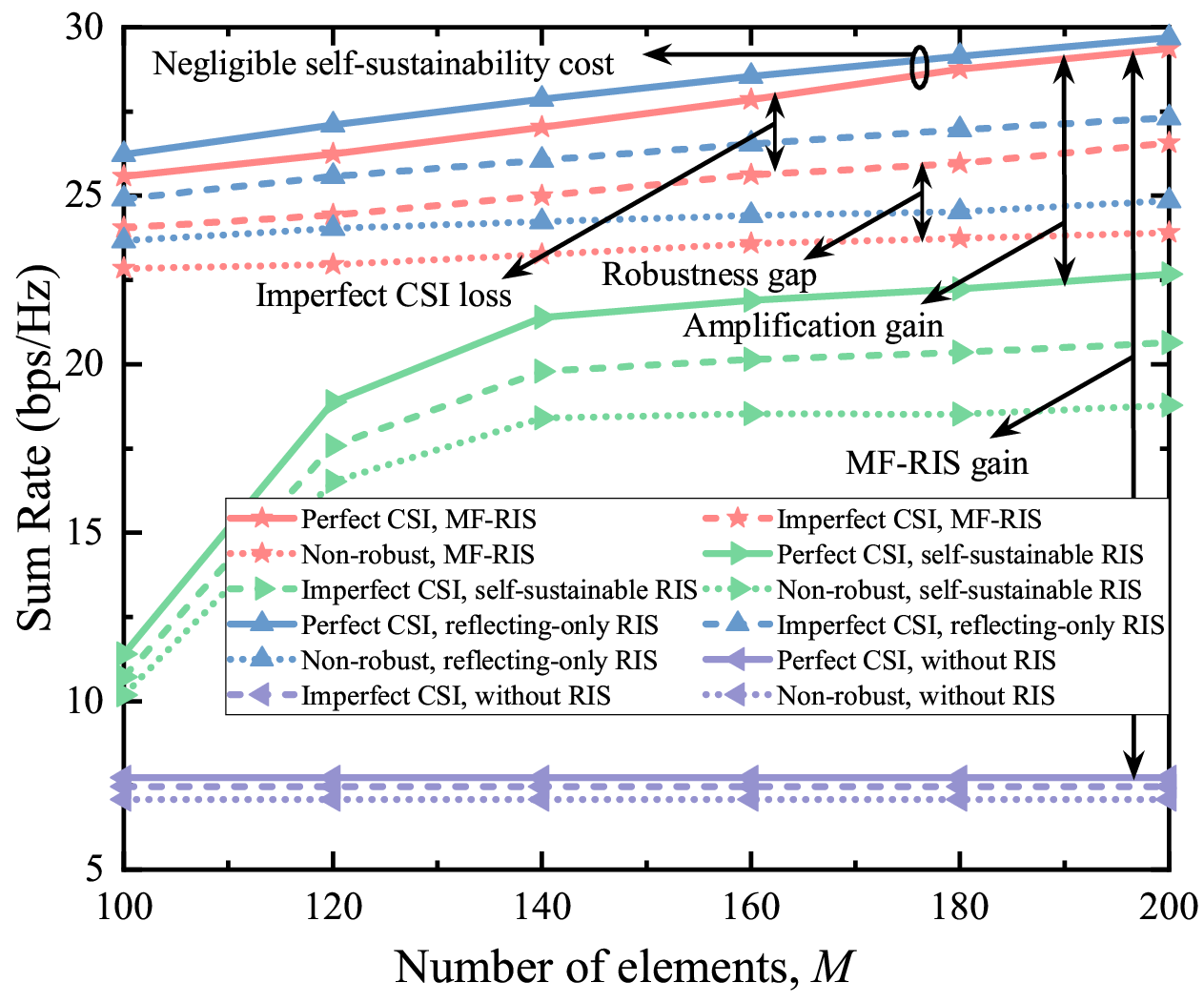}
    		\vspace{-2mm}
    		\caption{SR versus $M$ under different schemes and different CSI setups, where $P_{\rm BS}^{\max}=36$ dBm and $\beta_{\max}=16$ dB.}
    		\label{M}
    		\vspace{-1mm}
       \end{figure}
       \begin{figure}[t]
   	\centering
   	\vspace{-1mm}
   	\includegraphics[width=3.05in]{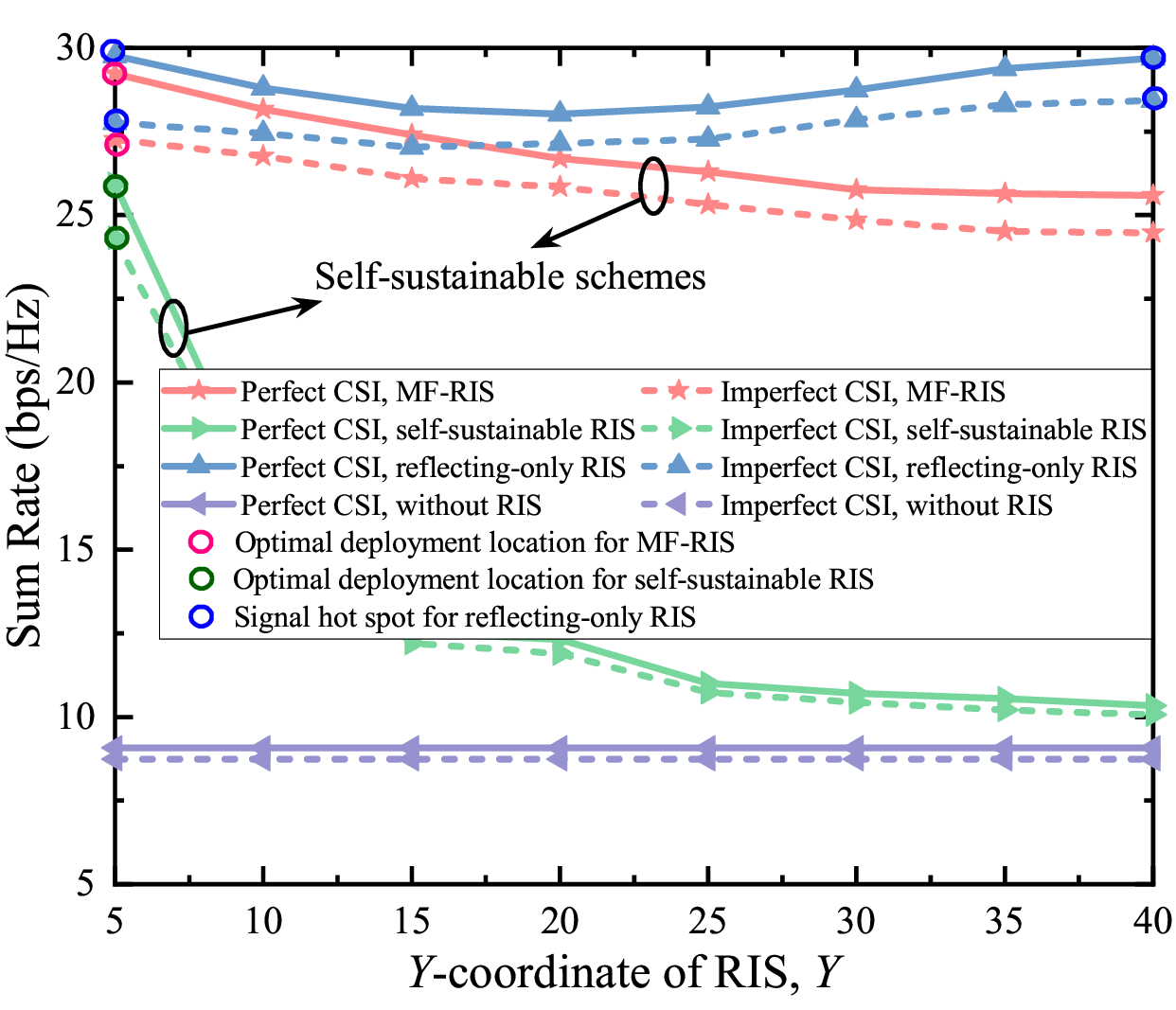}
   	\vspace{-2mm}
   	\caption{SR versus $Y$ under different schemes and different CSI setups, where $P_{\rm BS}^{\max}=40$ dBm, $M=120$, and $\beta_{\max}=16$ dB.}
   	\label{Y}
   	\vspace{-1mm}
   \end{figure}
    
    Fig. \ref{M} plots the achievable SR versus the number of RIS elements under different schemes and various CSI setups.
    ``Non-robust scheme" is the same as the proposed scheme except that it treats the estimated CSI (i.e., $\widetilde{\mathbf{h}}_k$, $\widetilde{\mathbf{g}}_k$, and $\widetilde{\mathbf{H}}$) as perfect CSI.
    First, the MF-RIS scheme shows impressive SR performance for different CSI uncertainties.
    In particular, when $M=200$, the MF-RIS schemes under perfect CSI, imperfect CSI, and non-robust cases attain 280\%, 256\%, and 238\% higher SR than the counterparts without RIS.
    Second, compared to the passive RIS, the performance loss of MF-RIS is negligible, especially for large-size surfaces.
    This behavior can be explained as follows:
    1) a larger-size MF-RIS can allocate more elements to operate in H mode, so that more energy can be collected to power the signal reflection and amplification circuits;
    and 2) a larger $M$ offers a higher beamforming flexibility, and thus effectively enhances the information transmission from the BS to all users.
    Third, it is observed from Fig. \ref{M} that the performance loss caused by the CSI uncertainty increases with $M$.
    This is because increasing $M$ results in a higher channel estimation error, which reduces the SR improvement.
    As such, we conjecture that within a reasonable region of CSI uncertainty, the benefits brought by the growth of $M$ can outweigh its drawbacks.
    However, if the CSI uncertainty exceeds the acceptable range, the SR gain would suffer a significant loss.
   
    Fig. \ref{Y} illustrates the impact of the distance between the BS and RIS on the SR performance by varying the $Y$-coordinate of RIS.
   We observe that in both imperfect and perfect CSI cases, when the reflecting-only RIS moves from the BS to users, the achievable SR first decreases and then increases.
    This is because the path loss is a decreasing function of distance.
    Reducing the distance between the BS and the RIS as well as the distance between the RIS and users increases the channel gains of RIS-aided cascaded links.
    Thus, the reflecting-only RIS should be deployed near the BS or users, as it can create signal hot spots for them.
    By contrast, the performance gains of the self-sustainable RIS and MF-RIS-aided schemes decrease as the RIS moves away from the BS, and the optimal value is obtained when the RIS is in close proximity to the BS.
    This can be explained as follows.
    As the distances of the BS-self-sustainable RIS and BS-MF-RIS links increase, the energy harvested by the RIS elements decreases.
    It can be observed from constraint (\ref{C_energy}) that in order to maintain energy self-sustainability, the self-sustainable RIS and MF-RIS have to allocate more elements for energy harvesting. 
    This results in fewer elements for signal reflection and amplification, which in turn affects the desired signal reception.
   
   \vspace{-2mm}
   \section{Conclusions}\label{Conclusion}
    This paper proposed a new MF-RIS architecture to address the double-fading attenuation and the grid/battery dependence issues faced by conventional passive RISs.
    By integrating signal reflection, amplification, and energy harvesting on one surface, the proposed MF-RIS is expected to achieve self-sustainability and an improved throughput.
    Based on the operating protocol of the proposed MF-RIS, we derived the achievable SNR for MF-RIS and self-sustainable RIS-aided systems to quantify the performance gain achieved by the MF-RIS.
    Next, we formulated SR maximization problems and provided efficient solutions for both perfect and imperfect CSI cases.
    Simulation results validated the effectiveness of the proposed MF-RIS to improve throughout performance in a self-sustainable manner.
    Experimental results also revealed the strong robustness of the proposed algorithm in terms of CSI imperfectness as well as the great ability to exploit large-size RISs.
    Furthermore, practical design guidelines for MF-RIS-assisted multi-user systems were provided.
    In particular, deploying MF-RIS close to the transmitter is favorable for harvesting more energy and reaping the throughput benefits offered by the MF-RIS.

   \vspace{-2mm}
    \appendices
	\section{Proof of Proposition \ref{proposition-1}} \label{proof_of_proposition_1}
  We first consider the case where constraint (\ref{analysis-problem-1-C1}) is active.
   According to the inequality $\beta_{\max}\leq \frac{P_{\rm O}^{\rm MF}(\boldsymbol{\alpha})}{ M_{\rm A}(P_{\rm BS}^{\max}h^2+\sigma_1^2)}$ and the definition $P_{\rm O}^{\rm MF}(\boldsymbol{\alpha})=\frac{1}{\xi}(\sum\nolimits_{m=1}^M P_m^{\rm A}-M_{\rm A}(P_b+P_{\rm DC})-M_{\rm H}P_{\rm C})$, we derive $M_{\rm A}\leq \frac{\sum\nolimits_{m=1}^M P_m^{\rm A}-M_{\rm H}P_{\rm C}}{ \xi \beta_{\rm max}(P_{\rm BS}^{\max}h^2+\sigma_1^2)+P_b+P_{\rm DC}}$ and $\beta^{\star}_m=\beta_{\rm max}$.
	Then, for the case where constraint (\ref{analysis-problem-1-C2}) is active, it is easy to obtain $M_{\rm A}> \frac{\sum\nolimits_{m=1}^M P_m^{\rm A}-M_{\rm H}P_{\rm C}}{ \xi \beta_{\rm max}(P_{\rm BS}^{\max}h^2+\sigma_1^2)+P_b+P_{\rm DC}}$ and $\beta^{\star}_m=\frac{P_{\rm O}^{\rm MF}(\boldsymbol{\alpha})}{ M_{\rm A}(P_{\rm BS}^{\max}h^2+\sigma_1^2)}$.
	
	\vspace{-2mm}
	\section{Proof of Proposition \ref{MF-RIS-SNR}} \label{proof_of_MF-RIS-SNR}	
      By substituting the optimal solutions (\ref{C_analysis-problem_solution1}) and (\ref{C_analysis-problem_solution2}) into the objective function (\ref{Function-SNR-analysis}), the maximum SNR in the MF-RIS-aided system is given by (\ref{Appendix-B-gamma_MF}) at the bottom of this page.
      \begin{figure*}[b]
      	\hrulefill
      	     \begin{eqnarray}
      	     	\label{Appendix-B-gamma_MF}
      		\gamma_{{\rm MF}}\!\!\!\!\!\!\!&&\!\!\!\!\!=\left\{ \begin{aligned}
      			&\frac{P_{\rm BS}^{\rm max}\beta_{\max}\big|\sum_{m=1}^M \alpha_{m} hg\big|^2}
      			{\sigma_1^2\beta_{\max}\sum_{m=1}^M\alpha_{m}^2g^2+\sigma_0^2}, & & M_{\rm A}\leq M_{\rm A,1},  \\ 
      			&\frac{P_{\rm BS}^{\rm max}P_{\rm O}^{\rm MF}(\boldsymbol{\alpha})\big|\sum_{m=1}^M\alpha_mhg\big|^2}
      			{P_{\rm O}^{\rm MF}(\boldsymbol{\alpha})\sigma_1^2\sum_{m=1}^M \alpha_{m}^2g^2+\sigma_0^2\sum_{m=1}^M\alpha_m^2(P_{\rm BS}^{\rm max}h^2+\sigma_1^2)}, & & M_{\rm A}> M_{\rm A,1}. \\ 
      		\end{aligned} \right.
      	\end{eqnarray}
      \end{figure*}
	Based on the definitions $M_{\rm A}=\sum_{m=1}^M \alpha_{m}$ and $M_{\rm H}=M-\sum_{m=1}^M \alpha_{m}$, and the mode indicator constraint $\alpha_{m}\in\{0,1\}$, the achievable SNR of the MF-RIS is further derived as (\ref{MF-RIS-SNR-2}). 
	
	\vspace{-2mm}
	\section{Proof of Proposition \ref{proposition-2}} \label{proof_of_proposition_2}	
	Denote $\gamma_{{\rm MF}}^{'}(M_{\rm A})$ as the first derivative of (\ref{MF-RIS-SNR-2}) with respect to $M_{\rm A}$, then it can be verified that for $M_{\rm A}\leq M_{\rm A,1}$, $\gamma_{{\rm MF}}^{'}(M_{\rm A})\geq 0$ always holds.
	While for the case of $M_{\rm A}> M_{\rm A,1}$, we deduce that when $M_{\rm A}\leq  M_{\rm A,2}$, $\gamma_{{\rm MF}}^{'}(M_{\rm A})\geq 0$, and when $M_{\rm A}> M_{\rm A,2}$, $\gamma_{{\rm MF}}^{'}(M_{\rm A})< 0$.
	Accordingly, for $M_{\rm A,2}\leq M_{\rm A,1}$, we have the following properties:
	1) when $M_{\rm A}\leq M_{\rm A,1}$,  $\gamma_{{\rm MF}}^{'}(M_{\rm A})\geq 0$ holds, and thus $\gamma_{{\rm MF}}(M_{\rm A})$ increases as $M_{\rm A}$ increases;
	and 2) when $M_{\rm A}> M_{\rm A,1}$, $\gamma_{{\rm MF}}^{'}(M_{\rm A})< 0$ holds, and thus $\gamma_{{\rm MF}}(M_{\rm A})$ decreases as $M_{\rm A}$ increases.
	Armed with 1) and 2), the optimal number of reflection elements is given by $M_{\rm A}^{\star}= M_{\rm A,1}$.
	Similarly, for $M_{\rm A,2}> M_{\rm A,1}$, we obtain that:
	1) when $M_{\rm A}\leq M_{\rm A,2}$,  $\gamma_{{\rm MF}}^{'}(M_{\rm A})\geq 0$ holds, and thus $\gamma_{{\rm MF}}(M_{\rm A})$ increases as $M_{\rm A}$ increases;
	and 2) when $M_{\rm A}<M_{\rm A,2} $, $\gamma_{{\rm MF}}^{'}(M_{\rm A})< 0$ holds, and thus $\gamma_{{\rm MF}}(M_{\rm A})$ decreases as $M_{\rm A}$ increases.
	Therefore, the optimal number of reflection elements is $M_{\rm A}^{\star}=M_{\rm A,2}$.
	Finally, the optimal number of reflection elements is obtained as (\ref{MF-Optimal-M}).

     \vspace{-2mm}
	\section{Proof of Proposition \ref{proposition-3}} \label{proof_of_proposition_3}	
	Similar to the proof of Proposition \ref{MF-RIS-SNR}, by substituting the optimal solutions (\ref{DF-RIS-solution-1}) and (\ref{DF-RIS-solution-2}) into (\ref{analysis-SE-function}), the achievable SNR of the self-sustainable RIS-aided system can be derived as (\ref{SNR-SE-RIS}).
	It can be observed that $\gamma_{\rm SE}(M_{\rm A})$ is an increasing function of $M_{\rm A}$.
	In addition, we can deduce from the energy constraint (\ref{power-DF-RIS}) that constraint $M_{\rm A}\leq \frac{\sum\nolimits_{m=1}^M P_m^{\rm A}-M_{\rm H}P_{\rm C}}{P_b}$ should be satisfied. 
	Thus, for this case, the optimal number of reflection elements is $M_{\rm A}^{\star}=\lfloor\frac{\sum\nolimits_{m=1}^M P_m^{\rm A}-M_{\rm H}P_{\rm C}}{P_b}\rfloor$.
	
	\vspace{-2mm}
	\section{Proof of Lemma \ref{Lemma-SCA}} \label{proof_of_Lemma_SCA}	
	Defining $x$ as a complex scalar variable and $\{x^{(\ell)}\}$ as a feasible point in the $\ell$-th iteration, then according to the FTS, we have the following inequality:
	\begin{eqnarray}
		\label{X_FTS}
		\left|x\right|^2\geq 2{\rm Re}\{(x^{(\ell)})^{\ast}x\}-(x^{(\ell)})^{\ast}x^{(\ell)}.
	\end{eqnarray}
	Next, by replacing  $x$ and $x^{(\ell)}$ in (\ref{X_FTS}) with $(\mathbf{h}_{k}^{\mathrm H}+\mathbf{v}^{\mathrm H} \mathbf{G}_{k})\mathbf{f}_{k}$ and $(\mathbf{h}_{k}^{\mathrm H}+(\mathbf{v}^{(\ell)})^{\mathrm H} \mathbf{G}_{k})\mathbf{f}_{k}^{(\ell)}$, respectively, a lower bound on the convex term $|\bar{\mathbf{h}}_{k} \mathbf{f}_{k}|^2$ is obtained as
	\begin{eqnarray}
	   \nonumber
		& 2 {\rm Re} \big\{\underbrace{(\mathbf{h}_{k}^{\mathrm H}+ (\mathbf{v}^{(\ell)})^{\mathrm H} \mathbf{G}_{k})\mathbf{f}_{k}^{(\ell)}\mathbf{f}_{k}^{\mathrm H} (\mathbf{h}_{k}+\mathbf{G}_{k}^{\mathrm H}\mathbf{v})}_{g_{1,k}}\big\} \\
	    \label{h_kf_k_lb}
		&~ -\underbrace{(\mathbf{h}_{k}^{\mathrm H}+(\mathbf{v}^{(\ell)})^{\mathrm H} \mathbf{G}_{k})\mathbf{f}_{k}^{(\ell)}(\mathbf{f}_{k}^{(\ell)})^{\mathrm H}(\mathbf{h}_{k}+\mathbf{G}_{k}^{\mathrm H}\mathbf{v}^{(\ell)})}_{g_{2,k}}.
	\end{eqnarray}
	Furthermore, by inserting $\mathbf{h}_{k}\!=\!\widetilde{\mathbf{h}}_{k} \!+\!\triangle\mathbf{h}_{k}$ and  $\mathbf{G}_{k}\!=\!\widetilde{\mathbf{G}}_{k} \!+\!\triangle\mathbf{G}_{k}$ into (\ref{h_kf_k_lb}) and performing mathematical transformations, we recast the first term in (\ref{h_kf_k_lb}), $g_{1,k}$, as (\ref{Appendix_g_1}) at the top of the next page.
	\begin{figure*}[t]
			\begin{eqnarray}
			\nonumber
			g_{1,k}~~\!\!\!\!\!\!\!\!\!\!\!\!\!\!&{}&=\big[(\widetilde{\mathbf{h}}_{k}^{\mathrm{H}}+\triangle\mathbf{h}_{k}^{\mathrm{H}})+ (\mathbf{v}^{(\ell)})^{\mathrm H}(\widetilde{\mathbf{G}}_{k}+\triangle\mathbf{G}_{k} )\big]\mathbf{f}_{k}^{(\ell)}\mathbf{f}_{k}^{\mathrm H} \big[(\widetilde{\mathbf{h}}_{k}+\triangle\mathbf{h}_{k})+(\widetilde{\mathbf{G}}_{k}^{\mathrm H}+\triangle\mathbf{G}_{k}^{\mathrm H})\mathbf{v}\big]\\
			\nonumber
			\!\!\!\!\!\!\!\!\!\!\!\!\!\!&{}&=(\widetilde{\mathbf{h}}_{k}^{\mathrm{H}}+(\mathbf{v}^{(\ell)})^{\mathrm H}\widetilde{\mathbf{G}}_{k})\mathbf{f}_{k}^{(\ell)}\mathbf{f}_{k}^{\mathrm H}
			(\widetilde{\mathbf{h}}_{k}+\widetilde{\mathbf{G}}_{k}^{\mathrm H}\mathbf{v})
			+(\widetilde{\mathbf{h}}_{k}^{\mathrm{H}}+(\mathbf{v}^{(\ell)})^{\mathrm H}\widetilde{\mathbf{G}}_{k})\mathbf{f}_{k}^{(\ell)}\mathbf{f}_{k}^{\mathrm H}
			(\triangle\mathbf{h}_{k}+\triangle\mathbf{G}_{k}^{\mathrm H}\mathbf{v})\\
			\nonumber
			\!\!\!\!\!\!\!\!\!\!\!\!\!\!&{}&~~+(\triangle\mathbf{h}_{k}^{\mathrm{H}}+(\mathbf{v}^{(\ell)})^{\mathrm H}\triangle\mathbf{G}_{k})\mathbf{f}_{k}^{(\ell)}\mathbf{f}_{k}^{\mathrm H}
			(\widetilde{\mathbf{h}}_{k}+\widetilde{\mathbf{G}}_{k}^{\mathrm H}\mathbf{v})
			+(\triangle\mathbf{h}_{k}^{\mathrm{H}}+(\mathbf{v}^{(\ell)})^{\mathrm H}\triangle\mathbf{G}_{k})\mathbf{f}_{k}^{(\ell)}\mathbf{f}_{k}^{\mathrm H}
			(\triangle\mathbf{h}_{k}+\triangle\mathbf{G}_{k}^{\mathrm H}\mathbf{v})\\
			\nonumber
			\!\!\!\!\!\!\!\!\!\!\!\!\!\!&{}&=(\widetilde{\mathbf{h}}_{k}^{\mathrm{H}}+(\mathbf{v}^{(\ell)})^{\mathrm H}\widetilde{\mathbf{G}}_{k})\mathbf{f}_{k}^{(\ell)}\mathbf{f}_{k}^{\mathrm H}
			(\widetilde{\mathbf{h}}_{k}+\widetilde{\mathbf{G}}_{k}^{\mathrm H}\mathbf{v})
			+(\widetilde{\mathbf{h}}_{k}^{\mathrm{H}}+(\mathbf{v}^{(\ell)})^{\mathrm H}\widetilde{\mathbf{G}}_{k})\mathbf{f}_{k}^{(\ell)}\mathbf{f}_{k}^{\mathrm H}\triangle\mathbf{h}_{k}\\
			\nonumber
			\!\!\!\!\!\!\!\!\!\!\!\!\!\!&{}&~~+{\rm vec}^{\mathrm H}{(\triangle\mathbf{G}_{k})}{\rm vec}(\mathbf{v}(\widetilde{\mathbf{h}}_{k}^{\mathrm{H}}+(\mathbf{v}^{(\ell)})^{\mathrm H}\widetilde{\mathbf{G}}_{k})\mathbf{f}_{k}^{(\ell)}\mathbf{f}_{k}^{\mathrm H})
			+\triangle\mathbf{h}_{k}^{\mathrm{H}}\mathbf{f}_{k}^{(\ell)}\mathbf{f}_{k}^{\mathrm H}
			(\widetilde{\mathbf{h}}_{k}+\widetilde{\mathbf{G}}_{k}^{\mathrm{H}}\mathbf{v})
			+\triangle\mathbf{h}_{k}^{\mathrm{H}}\mathbf{f}_{k}^{(\ell)}\mathbf{f}_{k}^{\mathrm H}\triangle\mathbf{h}_{k}\\
			\nonumber
			\!\!\!\!\!\!\!\!\!\!\!\!\!\!&{}&~~+{\rm vec}^{\mathrm H}(\mathbf{v}^{(\ell)}(\widetilde{\mathbf{h}}_{k}+\mathbf{v}^{\mathrm{H}}\widetilde{\mathbf{G}}_{k}^{\mathrm{H}})\mathbf{f}_{k}(\mathbf{f}_{k}^{(\ell)})^{\mathrm H}
			){\rm vec}(\triangle\mathbf{G}_{k})
			+{\rm vec}^{\mathrm H}(\triangle\mathbf{G}_{k})(\mathbf{f}_{k}^{\ast}(\mathbf{f}_{k}^{(\ell)})^{\mathrm T}\otimes \mathbf{v})\triangle\mathbf{h}_{k}^{\ast}\\
			\nonumber
			\!\!\!\!\!\!\!\!\!\!\!\!\!\!&{}&~~+\triangle\mathbf{h}_{k}^{\mathrm T}(\mathbf{f}_{k}^{\ast}(\mathbf{f}_{k}^{(\ell)})^{\mathrm T}\otimes (\mathbf{v}^{(\ell)})^{\mathrm H}{\rm vec}(\triangle\mathbf{G}_{k}) 
			+{\rm vec}^{\mathrm H}(\triangle\mathbf{G}_{k})( \mathbf{f}_{k}^{\ast}(\mathbf{f}_{k}^{(\ell)})^{\mathrm T}\otimes \mathbf{v}(\mathbf{v}^{(\ell)})^{\mathrm H}){\rm vec}(\triangle\mathbf{G}_{k})\\
			\label{Appendix_g_1}
			\!\!\!\!\!\!\!\!\!\!\!\!\!\!&{}&=\mathbf{x}_{k}^{\mathrm H}\mathbf{\widetilde{A}}_{k}{\mathbf{x}_{k}}
			+\mathbf{\widetilde{a}}_{k}^{\mathrm H}{\mathbf{x}_{k}}+\mathbf{x}_{k}^{\mathrm H}\mathbf{\widehat{a}}_{k}+\widetilde{a}_{k}, ~\forall k.
		\end{eqnarray}
	\hrulefill
	\end{figure*}
	Similarly, the second term in (\ref{h_kf_k_lb}), $g_{2,k}$, is rewritten as 
	\begin{eqnarray}
		\label{Appendix_g_2}
		g_{2,k}=\mathbf{x}_{k}^{\mathrm H}\mathbf{\widehat{A}}_{k}\mathbf{x}_{k}+\mathbf{\bar{a}}_{k}^{\mathrm H}\mathbf{x}_{k}+\mathbf{x}_{k}^{\mathrm H}\mathbf{\bar{a}}_{k}+\widehat{a}_{k}, ~\forall k,
	\end{eqnarray}
	where the introduced coefficients $\mathbf{\widetilde{A}}_{k}$, $\mathbf{\widehat{A}}_{k}$, $\mathbf{\widetilde{a}}_{k}$, $\mathbf{\widehat{a}}_{k}$, $\mathbf{\bar{a}}_{k}$, $\widetilde{a}_{k}$, and $\widehat{a}_{k}$ in (\ref{Appendix_g_1}) and (\ref{Appendix_g_2}) are given by (\ref{robust_A_SCA_coefficients}).
	According to (\ref{h_kf_k_lb})-(\ref{Appendix_g_2}), we finally obtain (\ref{robust_A_SCA}).

\end{document}